\documentclass[12pt,fleqn]{article}
\usepackage[verbose]{cite}
\usepackage{amssymb,epsfig}
\usepackage{times}
\newcommand{\newsection}[1]{\section{#1}\setcounter{equation}{0}}
\renewcommand{\thefootnote}{\fnsymbol{footnote}}

\def\be{\begin{equation}}
\def\ee{\end{equation}}
\def\bea{\begin{eqnarray}}
\def\eea{\end{eqnarray}}
\def\dis{\displaystyle}
\def\max{\mathrm{max}}

\def\pole{\mathrm{pole}}

\def\GeV{{\mathrm{GeV}}}
\def\MeV{{\mathrm{MeV}}}
\def\aem{\alpha}

\def\bm{\boldmath}
\def\branch{{\mathcal B}}
\def\cl#1{{#1\%\ \mathrm{C.L.}}}
\def\ea{\emph{et al.}}
\def\eff{\mathrm{eff}}
\def\eq#1{Eq.~(\ref{#1})}
\def\eqs#1#2{Eqs.~(\ref{#1})--(\ref{#2})}
\def\fig#1{Fig.~\ref{#1}}
\def\figs#1#2{Figs.~\ref{#1}--\ref{#2}}
\def\nnu{\nonumber}
\def\newp{\mathrm{NP}}
\def\Oi{{{\mathcal O}}}
\def\ol#1{\overline{#1}}

\def\sm{\mathrm{SM}}
\def\Re{{\mathrm{Re}}\,}
\def\rf{Ref.~\cite}
\def\rfs{Refs.~\cite}
\def\Sec#1{Sec.~\ref{#1}}
\def \a{\alpha}
\def \b{\beta}

\def \g{\gamma}

\def \m{\mu}
\def \n{\nu}

\def\euro#1#2#3{{Eur. Phys. J. C} {\bf #1}, #3 (#2)}

\def\ibid#1#2#3{{\it ibid.\/}~{\bf#1}, #3 (#2)}
\def\ib#1#2#3{{\bf#1},  #3  (#2)}
\def\jhep#1#2#3{{J.~High~Energy~Phys.}~{\bf #1}, #3 (#2)}

\def\np#1#2#3{{Nucl.~Phys.}~{\bf B#1}, #3 (#2)}

\def\pl#1#2#3{{Phys.~Lett. B}~{\bf #1}, #3 (#2)}

\def\prd#1#2#3{{Phys.~Rev. D}~{\bf #1}, #3 (#2)}
\def\prl#1#2#3{{Phys.~Rev.~Lett.}~{\bf #1}, #3 (#2)}

\def\rmp#1#2#3{{Rev. Mod. Phys.} {\bf #1}, #3 (#2)}
\def\zpc#1#2#3{{Z.~Phys. C}~{\bf #1}, #3 (#2)}
\begin{document}
\begin{titlepage}
\begin{flushright}
LMU 18/03\\
TUM-HEP-519/03 \\
hep-ph/0310219\\
October  2003
\end{flushright}
\vskip 0.2in
\begin{center}
\setlength{\baselineskip}{0.35in} 
{\bm\bf\Large More Model-Independent Analysis of \bm$b\to s$ Processes
}\\[1.2cm]
\setlength {\baselineskip}{0.2in}
{ \sc Gudrun Hiller\footnote{E-mail address: hiller@theorie.physik.uni-muenchen.de} }\\[1.3mm]
\emph{Ludwig-Maximilians-Universit\"at M\"unchen, Sektion Physik, Theresienstra\ss e 37, \\ D-80333 M\"unchen, Germany}
\vspace{2em}

{\sc Frank  Kr\"uger\footnote{E-mail address: fkrueger@ph.tum.de}}\\[1.3mm]
\emph{Physik Department, Technische Universit\"at M\"unchen,\\ D-85748 Garching, Germany}\\[0.5cm]
\end{center}

\begin{abstract}
We  study model-independently the implications of non-standard 
scalar and pseu\-do\-sca\-lar interactions for the decays $b\to s
\gamma$, $b\to s g$, $b\to s \ell^+\ell^-$ ($\ell=e,  \mu$) and $B_s\to \mu^+\mu^-$.
We find sizeable  renormalization effects from scalar and pseudoscalar
four-quark operators in  the radiative
decays and at $O(\alpha_s)$  in hadronic $b$ decays.~Constraints on the Wilson coefficients of an  extended operator basis are worked out.
Further, the ratios $R_H=\branch(B\to H \mu^+\mu^-)/\branch(B\to
H e^+ e^-)$, for $H=K^{(*)}, X_s$,  and
their  correlations with the $B_s\to \mu^+\mu^-$ decay are  investigated.
We show  that the Standard Model  prediction for these ratios defined
with the same cut on the dilepton mass for electron and muon modes, 
 $R_H= 1 + O(m^2_\mu/m^2_b)$,  has   a  much smaller theoretical 
uncertainty ($\lesssim 1\%$) than the one for the individual branching fractions. 
The present experimental limit $R_K\leqslant 1.2$  puts
constraints on scalar and pseudoscalar couplings, which are similar
to the ones from current data on $\branch(B_s\to \mu^+\mu^-)$.
We find that   new physics corrections to  $R_{K^*}$ and $R_{X_s}$
can reach $13 \%$ and $10\%$, respectively. 
\end{abstract}
\end{titlepage}
%
%
\renewcommand{\thefootnote}{\arabic{footnote}}
\setcounter{footnote}{0}

\newsection{Introduction}\label{intro}
Flavor-changing neutral currents
(FCNCs) are forbidden in the Standard Model (SM) 
at tree level and arise only at one loop. 
Hence, they are sensitive to quantum corrections from 
heavy degrees of freedom at and above the electroweak scale. 
The rare decays 
$b\to s \gamma$, $b\to s g$ and  
$b \to s \ell^+ \ell^-$, where $\ell = e$ or $\mu$,  are
such promising probes. Measurements of these processes are 
rapidly improving by the present generation
of $B$ experiments and in the not too distant future by the Tevatron 
and the LHC.
The analysis of $b\to s$ transitions can be systematically performed in terms of an effective low-energy theory
with the Hamiltonian (see, e.g., \rf{BBL})
\be\label{heff}
{\mathcal{H}}_{\eff}=-\frac{4 G_F}{\sqrt{2}} V_{tb}^{} V_{ts}^* \sum_i
[C_i (\mu) {\mathcal{O}}_{i}(\mu)  + C_i^\prime (\mu) {\mathcal{O}}_{i}^\prime (\mu)].
\ee
The operators
${\mathcal{O}}_{i}^{(\prime)}$ in \eq{heff} include dipole couplings
with a photon and a gluon and dilepton operators with vector and
axial-vector, as well as with scalar and pseudoscalar  Lorentz
structures.
They are given as\footnote{Our definition of  ${\mathcal O}_{S,P}$ is different from that of
\rfs{Logan:2000iv, Bobeth:2001sq,Bobeth:2001jm} (i.e., without the factor
of $m_b$) in order for $C_{S,P}$
to be dimensionless. As a consequence, the scalar and pseudoscalar operators 
have a non-vanishing anomalous dimension.} 
\bea\nnu
{\mathcal{O}}_{7} = \frac{e}{g_s ^2} m_b
(\bar{s} \sigma_{\mu \nu} P_R b) F^{\mu \nu}, \quad 
{\mathcal{O}}_{8}=\frac{1}{g_s} m_b
(\bar{s}_{\alpha} \sigma_{\mu \nu} T^a_{\alpha \beta} 
P_R b_{\beta})G^{a \mu \nu},
\eea
\bea\nnu
{\mathcal{O}}_{9} = \frac{e^2}{g_s^2} 
(\bar{s} \gamma_{\mu} P_L b)(\bar \ell \gamma^\mu \ell), \quad 
{\mathcal{O}}_{10}=\frac{e^2}{g_s^2}
(\bar{s}  \gamma_{\mu} P_L b)(  \bar \ell \gamma^\mu \gamma_5\ell),
\eea
\bea\label{eq:scalar}
{\mathcal{O}}_{S} =  \frac{e^2}{16 \pi^2} 
(\bar{s} P_R b ) ( \bar \ell  \ell), \quad 
{\mathcal{O}}_{P}=\frac{e^2}{16 \pi^2} 
(\bar{s}P_R  b)  (\bar \ell \gamma_5\ell).
\eea
The operators  ${\mathcal{O}}_{1-6}$ can be seen in \rf{NNLO:OB}.
The  primed operators in \eq{heff} can be obtained 
from their unprimed
counterparts by replacing $P_L \leftrightarrow P_R$.
In the SM as well as in models with minimal flavor violation (MFV)
where flavor violation is entirely ruled by the
CKM matrix, the Wilson coefficients
$C^{\prime}_i$ are suppressed by the strange quark Yukawa
coupling 
\bea\label{eq:SMCprime}
C^{\prime }_i \sim \frac{m_s}{m_b} C_i.
\eea
Furthermore, the SM contributions to scalar and pseudoscalar operators 
due to
neutral Higgs-boson exchange are tiny even
for taus since
\begin{eqnarray}
\label{eq:SMCI}
C^{\sm}_{S,P} \sim \frac{m_\ell m_b}{m_W^2}.
\end{eqnarray}
Thus, in the context of the SM only the operators  $\Oi_{7-10}$ 
matter for semileptonic and radiative $b \to s $ transitions.

Our plan is to  determine the coefficients
$C_i^{(\prime)}$ from a fit to the data and thereby testing the SM
\cite{Ali:1994bf}. 
At present the number of measured independent observables
is not sufficient, so one currently has to simplify the program and  deal 
with a restricted set of operators.
In this work we  analyze the decays $B \to X_s \gamma$, 
$B \to X_s \ell^+ \ell^-$, $B \to K^{(*)} \ell^+
\ell^-$, $B_s \to \mu^+ \mu^-$  
with the following assumptions:
\begin{itemize}
\item[(i)] The effects of  right-handed currents can be neglected, 
i.e., $C_i^\prime \simeq 0$.
\item[(ii)] The Wilson coefficients of scalar and pseudoscalar operators
are proportional to the lepton mass  $C_{S,P}\propto m_\ell$ such that  
the coupling to electrons is negligible. This is automatically
fulfilled if $C_{S,P}$ are generated by neutral Higgs-boson exchange, but
not in general within  SUSY models with broken $R$-parity.\footnote{Some
$R$-parity-violating SUSY models with horizontal flavor symmetries
do have $C_{S,P}\propto m_\ell$. They can generate in general
also helicity-flipped coefficients $C_i^\prime$ \cite{Guetta:1997fw}.}
\item[(iii)] There are no CP-violating phases from physics beyond the SM.
\end{itemize}

Therefore we take into account the Wilson
coefficients $C_{7-10}$ and $C_{S,P}$.
Model-independent analyses of  the decays $b \to s \gamma $ and $b
\to s \ell^+ \ell^-$
in the framework of the SM operator basis with ${\mathcal{O}}_{7-10}$ 
have been previously performed in 
\rfs{Hewett:1996ct,Ali:2002jg,Ali:1999mm}.
Distributions with an extended basis including ${\mathcal{O}}_{S,P}$ 
were  analyzed for $B \to X_s \ell^+ \ell^-$ in 
\rfs{Guetta:1997fw,Fukae:1998qy}
and for $B \to K^* \ell^+ \ell^-$ decays in \rfs{Aliev:1999gp,Yan:2000dc} 
to illustrate possible new physics 
effects. In these works, however,  no 
correlations between the
just-mentioned decay modes and $B_s\to \ell^+ \ell^-$
decays have been considered. 
In \rf{Bobeth:2001sq} the decays $B_s \to \mu^+ \mu^-$ and
$B \to K^{(*)} \mu^+ \mu^-$ 
have been studied model-independently.
It has been shown that the Wilson coefficients $C_{S,P}$ can be of $O(1)$
while respecting data on the $B_s \to \mu^+
\mu^-$ branching fraction, and thus are comparable in size to the 
vector and axial-vector couplings.
For a combined study of $B_s\to \mu^+ \mu^-$ and $B \to X_s \mu^+ \mu^-$ 
decays in the minimal supersymmetric standard model (MSSM), see \cite{Huang:2002ni}.

We perform here a combined analysis of the $B_s \to \mu^+ \mu^-$ 
branching ratio and the observables 
\begin{eqnarray}
\label{eq:RH}
R_{H}\equiv \frac{\dis \int_{4 m_\mu^2}^{q^2_{\max}}
dq^2\,  \frac{d \Gamma(B \to H  \mu^+ \mu^-)}{dq^2}}{\dis \int_{4
m_\mu^2}^{q^2_{\max}}
dq^2 \, \frac{d \Gamma(B \to H  e^+ e^-)}{dq^2}}, \quad H =
X_s, K^{(*)}, 
\end{eqnarray}
where $q^2_{\max}=(m_B-m_{K^{(*)}})^2$ for  $B \to K^{(*)} \ell^+ \ell^-$ and
$q^2_{\max}\approx m_b^{2}$ for
the inclusive decay modes. 
We also examine the low dilepton invariant mass region of the 
inclusive decays below the $J/\psi$ mass
with $q^2_{\max}= 6\ \GeV^2$.
Note that we use the lower cut of $4 m_\mu^2$ for both electron and muon
modes in order to remove phase space effects in the ratio $R_H$.
Within the SM, we obtain clean predictions even for  the exclusive 
decays
\begin{eqnarray}
R_{H}^{\sm}=1 + O ({m_\mu^2}/{m_b^2}),
\end{eqnarray}
which holds also outside the SM if $C_{S,P} \simeq 0$.
The normalization to the $e^+e^-$  mode
in \eq{eq:RH} was also discussed  in \rf{Wang:2003je} 
for the inclusive decays.

This paper is organized as follows.
In \Sec{exptl:status} we summarize the current experimental
status and constraints  on the decay modes of interest. 
Section \ref{NHB:four-quark:ops}
contains a discussion of new physics contributions to
scalar and pseudoscalar
four-quark operators and their impact on the Wilson coefficients of
the SM operator basis.~We investigate  new physics
effects in  the decays $b\to s \gamma$ and  $b\to s g$.
Model-independent constraints on the  coefficients of the operators $\Oi_{7-10}$
in the presence of $\Oi_S$ and $\Oi_P$ are
derived  in  \Sec{sec:constraints}.
In \Sec{correlations}
we study  correlations between the branching ratios of the decays $B_s \to\mu^+
\mu^-$, $ B \to X_s \ell^+ \ell^-$  and $B \to K^{(*)} \ell^+
\ell^-$. In particular,  quantitative predictions are obtained for the 
ratios $R_{K^{(*)}, X_s}$.
We summarize and conclude in \Sec{summary}. The anomalous dimensions,
decay distributions for $b\to s \ell^+\ell^-$ processes and auxiliary
coefficients are given in
Appendices \ref{tilde:basis}--\ref{recipe:Ai:Ci}.

\newsection{Experimental status of \bm $b\to s $ transitions }\label{exptl:status}

We summarize 
recent results on the inclusive and exclusive $b\to s
\ell^+\ell^-$  decay modes in Table
\ref{table:exptl:status:03}.
%
%
%
\begin{table}
\begin{center}
\caption{Branching fractions for various rare $B$
decays \cite{Aubert:2003rv,Aubert:2003cm,Abe:2003cp,Kaneko:2002mr}. 
The inclusive measurements as
well as the corresponding theoretical predictions  have
been obtained for $m_{e^+ e^-} > 0.2\ \GeV$. The SM predictions are
taken from \rf{Ali:2002jg} updated with $\branch(B\to X_c \ell
\nu_\ell)= 0.108$. \label{table:exptl:status:03} }
\smallskip

\begin{tabular}{lccc}\hline\hline
Decay modes & SM  & Belle  & BaBar \\
\hline \vspace{-1em}\\
$B\to X_s e^+e^-$& $(4.3\pm 0.7)\times 10^{-6} $& $(5.0\pm
2.3^{+1.3}_{-1.1})\times 10^{-6}$
&$(6.6\pm 1.9^{+1.9}_{-1.6})\times10^{-6} $ \\
$B\to X_s \m^+\m^-$&$(4.3\pm 0.7)\times 10^{-6} $ &$(7.9\pm
2.1^{+2.1}_{-1.5})\times 10^{-6}$&$(5.7\pm
2.8^{+1.7}_{-1.4})\times 10^{-6}$\\
$B\to Ke^+e^-$&$(3.6\pm 1.2)\times 10^{-7} $ &$(4.8^{+1.5}_{-1.3}\pm
0.3 \pm 0.1)\times 10^{-7} $
&$(7.4^{+1.8}_{-1.6}\pm 0.5)\times 10^{-7}$ \\
$B\to K\m^+\m^-$&$(3.6\pm 1.2)\times 10^{-7} $ &
$(4.8^{+1.2}_{-1.1}\pm 0.3 \pm 0.2)\times 10^{-7}$
&$(4.5^{+2.3}_{-1.9}\pm 0.4)\times 10^{-7} $  \\
$B\to K^*e^+e^-$& $(16.4\pm 5.1)\times 10^{-7}$ &
$(14.9^{+5.2}_{-4.6}{}^{+1.2}_{-1.3}\pm 0.2)\times
10^{-7}$&$(9.8^{+5.0}_{-4.2}\pm 1.1)\times 10^{-7} $\\
$B\to K^*\m^+\m^-$&$(12.4\pm 4.0)\times 10^{-7}$
&$(11.7^{+3.6}_{-3.1}\pm 0.9\pm 0.5)\times 10^{-7} $&$(12.7^{+7.6}_{-6.1}\pm 1.6)\times 10^{-7} $\\
\hline\hline\end{tabular}
\end{center}
\end{table}
%
%
These measurements are in agreement with
the SM prediction \cite{Ali:2002jg} within errors.
The experimental constraints we use in our
numerical calculations are given below.
Note that throughout this work we do not distinguish between $B$ and $\bar B$.

(i) The combined results of Belle \cite{Kaneko:2002mr} and 
BaBar \cite{Aubert:2003rv} for the inclusive $b\to s
\ell^+ \ell^-$ decays yield the $90\%$ confidence level intervals
\bea\label{eq:Xsee:90CL}
2.8\times 10^{-6}\leqslant \branch (B\to X_s e^+e^-) 
\leqslant 8.8\times 10^{-6},
\eea
\bea\label{eq:Xsmumu:90CL}
3.5\times 10^{-6}\leqslant \branch (B\to X_s \mu^+\mu^-) 
\leqslant 10.4\times 10^{-6}.
\eea
The statistical
significance of the Belle (BaBar) measurements of  
${\mathcal{B}}(B\to X_s e^+e^-)$ and
${\mathcal{B}}(B\to X_s \mu^+\mu^-)$ 
is $3.4\sigma$ ($4.0\sigma)$ and $4.7\sigma$ 
($2.2\sigma$), respectively.
To be conservative, we also use in our analysis 
the $\cl{90}$ limits \cite{Abe:2001qh}
\be\label{eq:belleXsee}
\branch (B\to X_s e^+e^-) < 10.1\times 10^{-6},
\ee
\be\label{eq:belleXsmu}
\branch (B\to X_s  \mu ^+\mu^-) < 19.1\times 10^{-6}
\ee
and compare their  implications with those of Eqs.~(\ref{eq:Xsee:90CL}) and
(\ref{eq:Xsmumu:90CL}).

(ii)  For the exclusive decay channels \cite{Aubert:2003cm, Abe:2003cp}
we obtain the following  $\cl{90}$
ranges  
\bea\label{eq:Kee:90CL}
3.9\times 10^{-7}\leqslant 
\branch (B\to K  e^+e^-) \leqslant 7.7\times 10^{-7},
\eea
\bea\label{eq:Kmumu:90CL}
3.0\times 10^{-7}\leqslant 
\branch (B\to K \mu^+\mu^-) \leqslant 6.5\times 10^{-7},
\eea
and
\bea\label{eq:Ksee:90CL}
6.5\times 10^{-7}\leqslant 
\branch (B\to K^* e^+e^-) \leqslant 17.9\times 10^{-7},
\eea
\bea\label{eq:Ksmumu:90CL}
6.7\times 10^{-7}\leqslant 
\branch (B\to K^* \mu^+\mu^-) \leqslant 17.0\times 10^{-7}.
\eea

(iii)  Using the experimental results displayed in 
Table \ref{table:exptl:status:03} we find for the ratios $R_H$ 
%
\bea
R_{X_s} = 1.20 \pm 0.52, \quad R_K = 0.81\pm 0.24,
\quad \left. R_{K^*}\right|_{\mathrm{no\ cut}} = 0.98\pm 0.38,
\eea
which translates into the $\cl{90}$ intervals
\be\label{ranges:90CL}
0.34 \leqslant R_{X_s} \leqslant 2.06, \quad 0.42 \leqslant R_K \leqslant
     1.20, \quad 0.35 \leqslant \left.R_{K^*}\right|_{\mathrm{no\ cut}}\leqslant 1.60.
\ee
Here, $\left. R_{K^*}\right|_{\mathrm{no\ cut}}$ is defined as $R_{K^*}$
with the lower integration boundary in the electron mode taken to be
$4 m_e^2$, since experimental data on the $B \to K^* e^+ e^-$ branching ratios are published
only for the  full phase space region. We do not include effects from
the small difference between the lower cut $m_{e^+e^-} = 0.2\  \GeV$ of
the experimental analysis \cite{Aubert:2003rv,Kaneko:2002mr} and $2 m_\mu$
used here. Furthermore, we neglect contributions to $B\to K
e^+e^-$ from the region below $q^2 = 4 m_\mu^2$, where the rate is tiny due
to the absence of the photon pole.
The above ratios should be compared with the predictions of the SM 
\bea
\label{eq:sm1}
R_{X_s}^\sm = 0.987\pm 0.006, \quad R^\sm_K = 1 \pm 0.0001,
\quad \left.R_{K^*}^{\sm}\right|_{\mathrm{no\ cut}}=0.73 \pm 0.01,
\eea
and 
\bea
\label{eq:sm2}
\left. R_{X_s}^\sm\right|_{\mathrm{low\ } q^2} = 0.977\pm 0.009,  
\quad R^\sm_{K^*} = 0.991 \pm 0.002,
\eea
where ``low $q^2$'' denotes a cut below 6 $\mbox{GeV}^2$.
The errors on the inclusive and exclusive ratios are due to a
variation of the renormalization scale and of the form factors,
respectively, see Secs.~\ref{sec:constraints} and \ref{correlations}.

(iv) The current world average of the inclusive $b\to s \gamma$ branching ratio
is \cite{bsgamma,exp:bsg}
\begin{eqnarray}\label{exp:bound:bsgamma}
{\mathcal{B}}(B \to X_s \gamma)=(3.34 \pm 0.38) \times 10^{-4}
\end{eqnarray}
with a photon energy cut $E_\gamma > m_b/20$.

(v) For the purely leptonic decays only upper limits exist. 
The branching ratio of the $B_s \to
\mu^+ \mu^-$ decay is constrained at $\cl{90}$ as \cite{exp:bmumu} 
\begin{eqnarray}
\label{eq:cdfbound}
{\mathcal{B}}( B_s \to \mu^+ \mu^-) < 2.0 \times 10^{-6}.
\end{eqnarray}
Note that there are preliminary ${90}\%$ confidence level limits of $9.5
\times 10^{-7}$  and $16 \times 10^{-7}$ from CDF and D\O,
respectively  \cite{nakao:LP:03}. 

\newsection{New physics contributions to four-quark operators}
\label{NHB:four-quark:ops}
In this section we address the question  whether  new physics contributions
to four-quark operators can spoil our model-independent analysis.
Firstly, the QCD penguins ${\mathcal{O}}_{3-6}$ appear in the SM and 
many extensions to lowest order only through operator mixing.
They enter the matrix element of $b \to s \gamma$ and 
$b \to s \ell^+ \ell^-$ decays at the loop level.
Hence, their impact is subdominant and new physics effects in QCD penguins
are negligible for our analysis within current precision.
Secondly, and this will be the important effect discussed 
in  the remainder of this section, 
it is conceivable that the dynamics which generates large
couplings to dileptons, i.e., to the operators ${\mathcal{O}}_{S,P}$, 
induces contributions to 4-Fermi operators 
with diquarks as well.  We introduce the following fermion  $f$ dependent operators
\begin{eqnarray}\label{new:ops:scalar}
{\mathcal{O}}_{L}^f =  (\bar{s} P_R b)(
\bar{f} P_L  f), \quad
{\mathcal{O}}_{R}^f= (\bar{s} P_R  b)(
\bar{f} P_R f),
\end{eqnarray}
where for muons we identify the coefficients $C_{L, R}^\mu = e^2/(16 \pi^2) (C_S \mp C_P)$.
We generalize here our assumption  (ii) in the sense that the coupling 
strength is proportional to the {\it fermion} mass $m_f$, 
which naturally arises in models with Higgs-boson  exchange.
In particular, the corresponding Wilson coefficients for $b$ quarks 
proportional to $m_b$ can be potentially large. 
%
%
As will be discussed in the next section,
current experimental data on the branching fraction 
of $B_s\to\mu^+\mu^-$ imply\footnote{In the MSSM with large $\tan \beta$ there are corrections to
the down-type Yukawa coupling
(see e.g.~\rfs{SUSY:large:tanbeta:Yukawa,Buras:2002wq}). 
These corrections can be substantial in $B$ decays, and have 
the form $1/(1+\epsilon_b\tan\beta)$ with $|\epsilon_b|\lesssim 0.01$ \cite{Buras:2002wq}.}
\begin{eqnarray}
\label{eq:clbbound}
\sqrt{ |C_{L}^b(m_W)|^2+|C_{R}^b(m_W)|^2 } \leqslant  
\frac{e^2}{16 \pi^2} \frac{{m}_b(m_W)}{m_\mu} 
\sqrt{2(|C_{S}(m_W)|^2+|C_{P}(m_W)|^2)} \lesssim
0.06.
\end{eqnarray}
Here, we anticipated our result in Eq.~(\ref{eq:CSPbound}), i.e., an 
upper bound on $|C_{S,P}|$ and evolved according to
$ {\rm d} (C_{S, P}(\mu)/ m_b (\mu)) {\rm d} \mu=0$, with the running
$b$-quark mass in the $\overline{\mathrm{MS}}$ scheme given in \eq{eq:mbrunning}.

The Wilson coefficients $C_{L, R}^b$  are non-zero to lowest order 
interactions at the electroweak scale $\mu \sim m_W$ and
can be significantly  larger than the ones of the 
QCD penguins $C_{3-6}(m_b)\sim O(10^{-2})$.
Hence, we have to study the potential impact of the operators 
${\mathcal{O}}_{L,R}^f$ on our analysis of $b \to s \gamma$ and
$b \to s \ell^+ \ell^-$ decays.

\subsection{One-loop mixing with pseudoscalar and  scalar operators}\label{sec:III:sub}

\begin{figure}
\begin{center}
\includegraphics[scale=0.55]{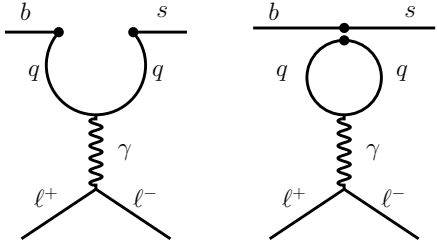}
\end{center}
\caption[]{Diagrams with an insertion of four-quark operators which 
contribute to the renormalization and the matrix element of the
operator $\widetilde{\mathcal{O}}_9$, and with an
on-shell photon and no leptons to $\widetilde{\mathcal{O}}_7$.}
\label{fig:loop}
\end{figure}

Scalar and pseudoscalar four-quark operators enter
radiative and semileptonic rare $b \to s $ decays at one-loop level as 
shown in \fig{fig:loop}.
To estimate their impact, we
insert  ${\mathcal{O}}_{L}^b$ and  ${\mathcal{O}}_{R}^b$
into the penguin diagrams with an internal $b$ quark
and use fully anticommuting $\gamma_5$.
%
%
%
%
The contributions from the diagram with
closed fermion loop vanish by Dirac trace and by gauge invariance or 
vector current conservation, i.e., after contraction with the lepton
current. 
For simplicity, we work in the ``standard'' operator
basis $\widetilde{\mathcal{O}}_{i}$ given in Appendix \ref{tilde:basis}. 
We obtain non-vanishing contributions from ${\mathcal{O}}_{R}^b$ and 
${\mathcal{O}}_{L}^b$ to $\widetilde{\mathcal{O}}_7$ and 
$\widetilde{\mathcal{O}}_9$, respectively.
The  diagrams with an internal $s$ quark contribute
to the helicity-flipped coefficients. They are
suppressed by a factor $m_s/m_b$ and therefore can be neglected.
We obtain the following corrections to the Wilson coefficients
at the scale $\mu_b=m_b$
\bea\label{eq:deltac7}
\delta \widetilde{C}_7(m_b)  =  \frac{1}{6} 
\ln \frac{m_W^2}{m_b^2} C_R^b(m_W), 
\eea
\be\label{eq:deltac9}
\delta \widetilde{C}_9(m_b)  =  \frac{1}{9} 
\ln \frac{m_W^2}{m_b^2} C_L^b(m_W).
\ee
These infinite renormalization contributions survive in the limit
 $\alpha_s \to 0$, which 
is similar to what happens in the SM for the mixing of
$\widetilde{\mathcal{O}}_2$ onto    $\widetilde{\mathcal{O}}_9$
\cite{Grinstein:MB}.
With the upper bound in
\eq{eq:clbbound} we find that 
the new physics effect from ${\mathcal{O}}_{L,R}^b$ 
is  
small, of the order of one percent for $\widetilde {\mathcal{O}}_9$, but 
$\mbox{a\  few} \times {\mathcal{O}}(10 \%)$ for  
$\widetilde {\mathcal{O}}_7$. 
The reason is simply that  $\widetilde C_7^{\sm}(m_b)$
is more than an order of magnitude smaller than  $\tilde
C_9^{\sm}(m_b)$, which in addition has a smaller  anomalous dimension.

Other operators contributing in the SM but  subleading 
in the decays $b \to s \gamma$ and $b\to s \ell^+\ell^-$ are also subject to
similar new physics effects.
To be specific, the Wilson coefficients of
the chromomagnetic dipole operator 
and the QCD penguin operators   receive
corrections  from the diagrams in \fig{fig:loop} with
diquarks instead of leptons  and  the intermediate 
photon replaced by a gluon. We find 
\be\label{eq:deltac8}
\delta \widetilde{C}_8(m_b) =  -\frac{1}{2} \ln \frac{m_W^2}{m_b^2} C_R^b(m_W),
\ee
\be\label{eq:deltac3}
\delta \widetilde{C}_{3,5}(m_b) =  -\frac{1}{18}  \frac{\alpha_s}{4 \pi}
\ln \frac{m_W^2}{m_b^2} C_L^b(m_W),
\ee
\be\label{eq:deltac4}
\delta \widetilde{C}_{4,6}(m_b) =  \frac{1}{6}  \frac{\alpha_s}{4 \pi}
\ln \frac{m_W^2}{m_b^2} C_L^b(m_W),
\ee
which are relevant to hadronic $B$ decays.\footnote{The decay $B \to \phi K_S$ has been
studied in \rf{Cheng:2003im} including $O(\alpha_s)$
corrections to the matrix element. The leading logarithmic contributions in
\eqs{eq:deltac8}{eq:deltac4}, however, have not been
taking into account, which explains the huge $\mu$ dependence 
found in these papers. We 
checked that the $\ln( m_b/\mu)$ terms of the $O(\alpha_s)$
corrections are canceled by the contributions in 
\eqs{eq:deltac8}{eq:deltac4}.} 
Quantitatively, the renormalization of the gluon dipole operator 
can be order one. (We study this in more detail below.)
The impact on the QCD penguins can be up to several percent.
As mentioned earlier, new physics contributions to the
operators $\widetilde{\mathcal{O}}_{3-6}$ are  subdominant  in 
$b \to s \gamma$ and $b\to s\ell^+ \ell^-$ 
decays. Since the
renormalization of $\widetilde{\mathcal{O}}_{9}$ by scalar and pseudoscalar
operators is small, too, 
we can safely neglect the effects of induced four-quark operators
of the type ${\mathcal{O}}_{L}^b$ in our analysis of semileptonic
and radiative $b \to s$ decays. 
We remark that scalar and pseudoscalar
operators also mix with the
electroweak penguin operators 
$\widetilde {\mathcal{O}}^e_{7-10}$  (see Appendix
   \ref{tilde:basis}) at order $\alpha/4 \pi$. We have calculated for 
completeness  the corresponding anomalous dimensions,
which can be seen in Appendix \ref{sec:mixing:ops}.

To get a more accurate estimate of the new physics corrections to
the magnetic penguin operators, we resum the leading logarithms in 
Eqs.~(\ref{eq:deltac7}) and 
(\ref{eq:deltac8}) by means of the renormalization group equations in
the $\overline{\mathrm{MS}}$ scheme \cite{BBL}.
Both operators ${\mathcal{O}}_{R}^b$ and ${\mathcal{O}}_{L}^b$ induce
additional operators under renormalization (see Appendix \ref{sec:mixing:ops}).
The anomalous dimensions of each set are known at 
next-to-leading order (NLO) \cite{Buras:2000if}, with no mixing between 
the sets.
We have calculated the  leading-order mixing of
${\mathcal{O}}_{L,R}^b$ onto
$\widetilde{\mathcal{O}}_{3-9}$.\footnote{The computation of the
anomalous dimensions at NLO is being performed in
\rf{CB:TE}.} The  
anomalous dimensions
are given in Appendix \ref{sec:mixing:ops}
together with the respective 
leading-order self-mixing of both ${\mathcal{O}}_{R}^b$
and ${\mathcal{O}}_{L}^b$ sets.
Numerically, we obtain
\bea\label{contr:CRb}
\delta \widetilde{C}_7(m_b)\simeq 0.71 C_R^b(m_W), \quad
\delta \widetilde{C}_8(m_b)\simeq -2.95 C_R^b(m_W),
\eea
which implies sizeable contributions to the branching ratios of
the radiative decays. We study the phenomenology in \Sec{sec:a7a8}.

The  mixing of scalar and
pseudoscalar operators in \eq{add:operators} onto the dipole operators
has been studied previously in the context of the two-Higgs-doublet
model \cite{Dai:etal} and in supersymmetry with gluino
contributions to $b \to s \gamma$\cite{Borzumati:1999qt}. While our
results
agree with the ones  presented in  \rf{Borzumati:1999qt}, 
they are at variance with those given in \rf{Dai:etal}. In particular, we disagree with the conclusion made
therein that the scalar and pseudoscalar operators do not mix 
with $\widetilde{\mathcal O}_9$.

\subsection{Implications for the decays \bm$b\to s \gamma$ and
$b\to s g$ \label{sec:a7a8}}
\begin{figure}
\begin{center}
\includegraphics[scale=0.55]{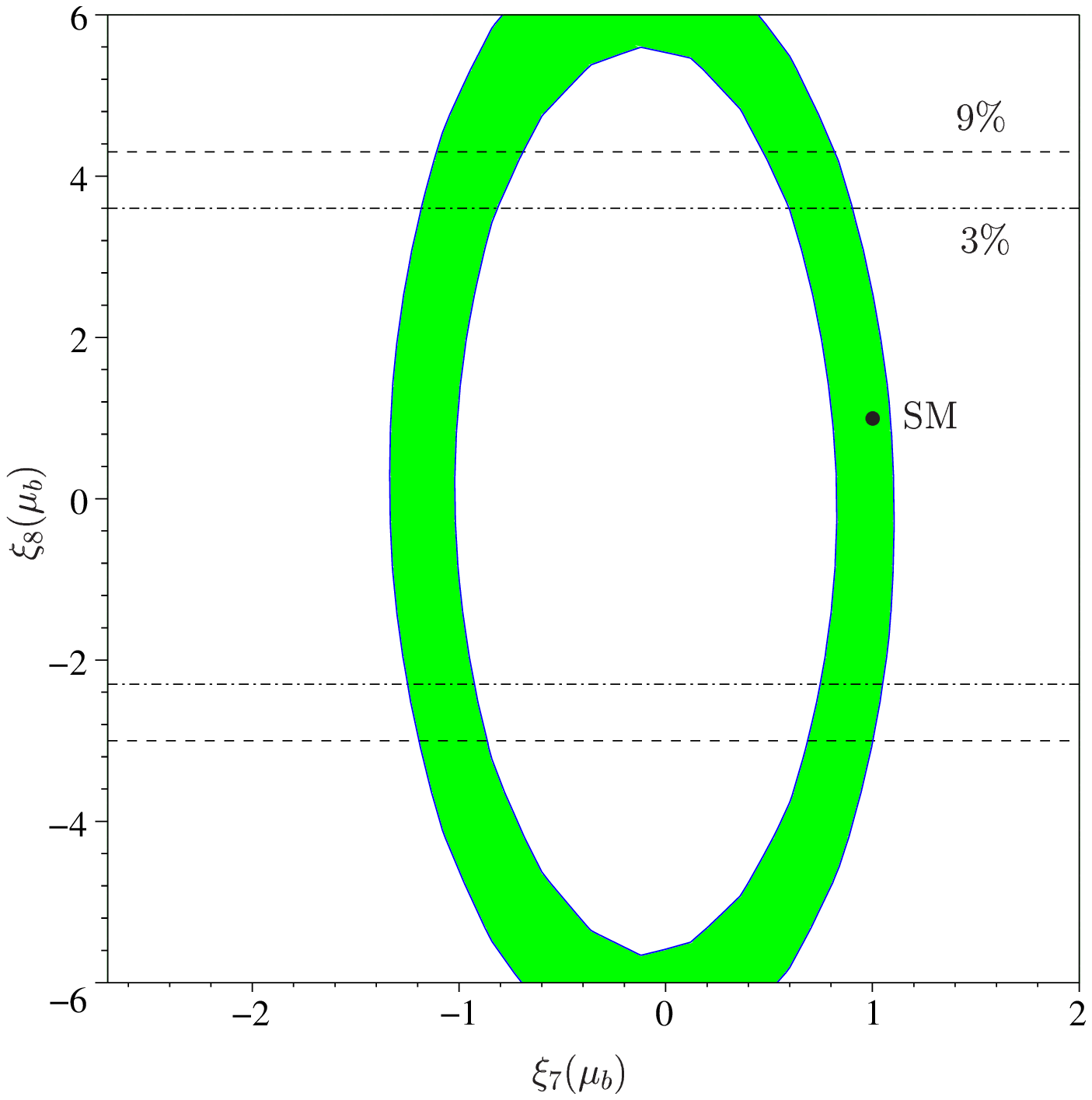}
\hspace{1em}
\includegraphics[scale=0.55]{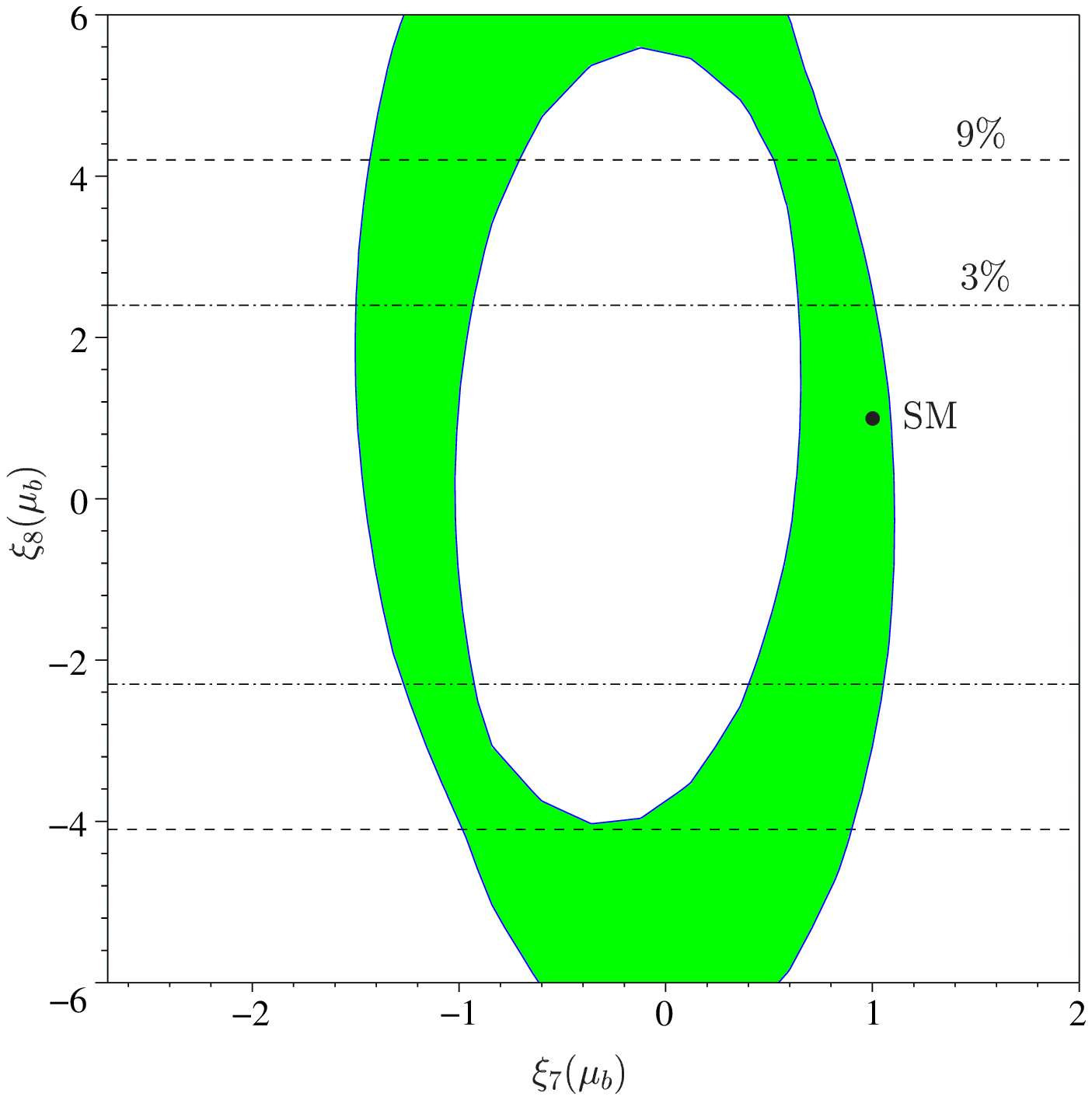}
\end{center}
\caption[]{Constraints on $\xi_{7,8}(\mu_b)$  from 
${\mathcal{B}}(b\to s \gamma) $ 
for $C_R^b(M_W)=0$ (left plot) and $C_R^b(M_W)=0.06$ (right plot).
We also show the upper and lower bounds on $\xi_8(\mu_b)$ for
the experimental limit $\branch(B\to X_s g) < 9\%$ 
\cite{cleo:bsg} (dashed lines)
and for an assumed value of $\branch(B\to X_s g) <3\%$ (dash-dotted lines).
\label{fig:bsgamma}}
\end{figure}

We now investigate the phenomenological consequences of the
mixing effects presented above for radiative $B$ decays. To illustrate how
large these corrections can be,
we normalize the Wilson coefficients in the 
presence of new physics to the ones in the SM, 
and denote this ratio by  $\xi$,  such that $\xi^\sm=1$. We obtain  to
next-to-leading order in the SM operator  basis and to leading logarithmic 
approximation in $C_R^b$
\bea
\xi_7(m_b) =  0.514 + 0.450\, \xi_7(m_W) + 0.035\, \xi_8(m_W)  -  2.319 \,C_R^b (m_W),
\eea
\bea
\xi_8(m_b)=  0.542 + 0.458\, \xi_8(m_W)  +19.790\,C_R^b (m_W) .
\eea
Given the upper bound  in \eq{eq:clbbound}
corrections of up to $14 \% $  and $119 \% $ to $\xi_7$ and $\xi_8$
can arise. 
We work out correlations between $\xi_7$ and $\xi_8$ 
from  ${\mathcal{B}}(B\to X_s \gamma)$ given in \eq{exp:bound:bsgamma}
and ${\mathcal{B}}(B\to X_s g)<9 \%$ at $90 \%$ C.L.~\cite{cleo:bsg},
using the analytical formulae of  \rfs{chetyrkin,Kagan:1998ym, greub:bsg}. 
We obtain the allowed regions at the $\mu_b$ scale shown in 
\fig{fig:bsgamma} for
$C_R^b (m_W)=0$ (left plot) and $C_R^b (m_W)=0.06$ (right plot).  
The theoretical uncertainty from the prescription of the charm-quark mass 
has been taken into account by including both solutions obtained for  
$m_c/m_b=0.22$  and $0.29$ \cite{GM}. 
{}From \fig{fig:bsgamma} we see that  
$A_7=0$  for  $C_R^b (m_W)=0.06$ is allowed by present data on the $b\to s g$ branching fraction. 
This particular scenario  could    be excluded by an improved experimental
analysis of $b\to sg$.
Also, if $C_R^b(m_W)$ is near its upper
bound, it implies  a contribution
to the matching conditions for $\widetilde{C}_{7,8}(m_W)$ in order to be consistent with experimental
data.

%
%
%
%

In summary, we find that the impact of $C_L^b$ on the matrix element of
$\widetilde{\mathcal{O}}_9$
is small, at most a few percent, and thus can be neglected. On
the other hand, contributions to the dipole operators
are in general 
non-negligible.
They can be avoided assuming $C_R^b (m_W)\simeq  0$, i.e.,
\bea
\label{eq:CsCpnoRH}
C_S+C_P=0.
\eea
In the remainder of this work 
we  discuss the phenomenology with and without this constraint.
Note that the absence of logarithms in the matching conditions for
$\widetilde C_{7,8}(m_W)$  from neutral Higgs-boson  exchange in a
two-Higgs-doublet model type II \cite{DAmbrosio:2002ex, Buras:2002wq}
is consistent with the fact that in this model
Eq.~(\ref{eq:CsCpnoRH}) is satisfied
\cite{Logan:2000iv}. 
This is also the case for the MSSM with MFV at large
$\tan \beta$ \cite{Bobeth:2001sq}.

\newsection{Model-independent analysis}\label{sec:constraints}

In this section we give the theoretical framework that we use to analyze
the decays $B \to X_s \gamma$, 
$B \to X_s \ell^+ \ell^-$, $B \to K^{(*)} \ell^+ \ell^-$, 
$B_s \to \mu^+ \mu^-$.
We then work out model-independent constraints on
the coefficients of the operators ${\mathcal{O}}_{7-10}$ and 
${\mathcal{O}}_{S,P}$. 

\subsection{Wilson coefficients and matrix elements}

The matrix element of inclusive $b \to s \ell^+ \ell^-$ decays 
contains contributions from 
the photon dipole operator ${\mathcal{O}}_7$, the dilepton operators
$\Oi_{9,10}$ and in models beyond the SM  also from $\Oi_{S,P}$.
The decay distributions in the SM are known to next-to-next-to-leading
order (NNLO) \cite{NNLO:OB,NNLO,Ghinculov:2002pe,Gambino:2003zm}, which corresponds to
NLO in $b \to s \gamma$. 
We use the NNLO expressions for the operators ${\mathcal{O}}_{7,9,10}$
and lowest order ones for ${\mathcal{O}}_{S,P}$
since $O(\alpha_s)$ corrections to the matrix elements of  
leptonic scalar and pseudoscalar
operators in these decays are not known.

Further, we assume that the contribution 
from intermediate charmonia has been removed
with experimental cuts.
Non-perturbative corrections 
\cite{non-pert} affect the branching 
ratio by at most few percent and we do not consider them here.
We neglect the mass of the strange quark but keep 
the muon mass consistently, because according to our assumption  (ii)
also $C_{S,P}$ counts as one power of $m_l$ 
and can be enhanced in models beyond the SM.

The dilepton invariant mass spectra  for inclusive and exclusive 
$b\to s \ell^+ \ell^-$ decays are given in Appendix \ref{dist}.
The effective coefficients which  enter the decay distributions
are written as  \cite{NNLO:OB,NNLO}
\be
\label{eq:c7tilde}
\widetilde{C}_7^{\eff}=
\Bigg[1+ \frac{\alpha_s(\mu)}{\pi} \omega_7(\hat s) \Bigg] A_7(\mu)-
\frac{\alpha_s(\mu)}{4\pi} \Bigg[\sum_{i=1,2} F^{(7)}_{i}(\hat
s)C_i^{(0)}(\mu) +  F^{(7)}_{8}(\hat s)A_8^{(0)}(\mu)\Bigg], 
\ee
\bea
\widetilde{C}_9^{\eff} &=&
\Bigg[1+ \frac{\alpha_s(\mu)}{\pi} \omega_9(\hat s) \Bigg] [A_9(\mu)+
T_9  h (\hat m_c^2, \hat{s})+U_9  h (1,\hat{s})+
    W_9  h (0,\hat{s})]\nnu\\
&-& \frac{\alpha_s(\mu)}{4\pi}\Bigg[\sum_{i=1,2} F^{(9)}_{i}(\hat s)
  C_i^{(0)}(\mu)+  F^{(9)}_{8}(\hat s)A_8^{(0)}(\mu)\Bigg],
\label{eq:c9tilde}
\eea
\be
\widetilde{C}_{10}^{\eff}=
\Bigg[1+ \frac{\alpha_s(\mu)}{\pi} \omega_9(\hat s) \Bigg] A_{10}(\mu),
\label{eq:cis:NNLO}
\ee
where $\hat{m}_c= m_c/m_b$, $\hat{s}=q^2/m_b^2$ and $A_i, T_9,U_9,W_9$
are given  in Appendix \ref{recipe:Ai:Ci}.  
The function $h(z, \hat{s})$ originates from the one-loop matrix elements 
of the four-quark operators  ${\mathcal{O}}_{1-6}$ (see
\fig{fig:loop}) and can be found in \rf{NNLO:OB}.
The functions
$\omega_i, F_{ij}$ arise from real and virtual  $\alpha_s$
corrections. They can be seen in \rfs{NNLO:OB,NNLO}
together with $\omega_{79}$ which replaces 
$\omega_7$ and  $\omega_9$ in the interference term
$\Re(\widetilde{C}_{7}^{\eff} \widetilde{C}_{9}^{\eff \ast})$ in the
decay rate. 
In the calculation of the  decay rate we expand in  powers of 
$\alpha_s$ and retain only linear terms.
Note that the $\omega_i$ include only that part from real
gluon emission which is required to cancel the divergence from the virtual 
corrections to the matrix element of the ${\mathcal{O}}_i$. Further
gluon bremsstrahlung corrections in $b \to s \ell^+ \ell^-$ 
decays \cite{Asatryan:2002iy, Ghinculov:2002pe} are subdominant over the whole 
phase space except for very low dilepton mass 
and are not taken into account here.
In our numerical analysis we choose a low value for the 
renormalization scale, $\mu_b= 2.5\ \GeV$, because
this approximates the full NNLO dilepton spectrum 
by the partial one, i.e., with the
virtual $O(\alpha_s)$ corrections $F_{1,2,8}^{(7,9)}=0$ in 
Eqs.~(\ref{eq:c7tilde}) and (\ref{eq:c9tilde})
switched off \cite{Ali:2002jg}.
This is beneficial since the $ F_{ij}$ are known in a compact analytical form
only for the low dilepton invariant mass region  \cite{NNLO}.
For the exclusive $B\to K^{(*)} \ell^+\ell^-$ decays we set 
$\omega_i=0$, since these corrections are already included
in the corresponding form factors. We do not take into account
hard spectator interactions \cite{beneke:etal}.  

Below  we work out model-independent bounds on  
$A_i\equiv A_i^{\sm}+A_i^{\newp}$. They differ from the ``true'' Wilson coefficients 
$C_{i}$ by penguin contributions that restore the 
renormalization scheme independence of the matrix element
\cite{Grinstein:MB}. 
In addition $A_9$ contains logarithms from insertions of the 
four-quark operators 
${\mathcal{O}}_{1 - 6}$ into the diagrams of \fig{fig:loop}.
Explicit formulae relating $A_i$ and $C_i$ 
are given in Appendix \ref{recipe:Ai:Ci}.
As discussed in \Sec{NHB:four-quark:ops}, we neglect
new physics contributions to the QCD penguin operators.
In our numerical study we use $f_{B_s}= 200\ \MeV$ and  $238\ \MeV$ \cite{Becirevic:2002zp}
and the parameters given in Table II  of \rf{Ali:2002jg} except for
$\branch(B\to X_c \ell  \n_\ell) = 10.80\%$ 
\cite{HFAG}. Form factors and their variation are taken from \rf{Ali:1999mm}.
We give the SM values for completeness:    $A_7^{\sm}(2.5\  \GeV)=-0.330$, 
$A_9^{\sm}(2.5\ \GeV)=4.069$ and
$A_{10}^{\sm}=-4.213$.

\subsection{Constraints from \bm $B_s\to \mu^+\mu^-$}

An upper limit on the $B_s\to \mu^+\mu^-$  branching ratio 
constrains the scalar and pseudoscalar couplings 
\bea\label{eq:CSPbound}
\sqrt{|C_S(\mu)|^2+|C_P(\mu)+\delta_{10}(\mu)|^2 } &\leqslant& 3.3 
\Bigg[\frac{{\mathcal{B}}( B_s \to \mu^+ \mu^-)}{2.0 \times
10^{-6}}\Bigg]^{1/2}\nnu\\
&\times&
\Bigg[\frac{|V_{tb}^{}V^*_{ts}|}{0.04}\Bigg]^2
\Bigg[\frac{{m}_b(\mu)}{4.4\ \GeV}\Bigg]
\Bigg[\frac{238\ \MeV}{f_{B_s}}\Bigg] \Bigg[\frac{1/133}{\aem}\Bigg].
\eea
Here, we neglected  the factor $(1- 4{m}_\mu^2/m_{B_s}^2 )$ in front of
$|C_S|^2$, see \eq{eq:brbmm}, and defined
$\delta_{10}(\mu) =2 m_\mu {m}_b(\mu) /m_{B_s}^2  A_{10}$.
The bound given in \eq{eq:cdfbound} also implies the upper limits
\bea
{\mathcal{B}}(B_s \to e^+ e^-) \leqslant  4.7 \times 10^{-11}, \quad  
{\mathcal{B}}(B_s \to \tau^+ \tau^-) \leqslant  4.2 \times 10^{-4}.
\eea

\subsection{Constraints from \bm$b\to s \gamma$}

The measured $b\to s \gamma$ branching fraction puts constraints on the
dipole operators.
In the absence of scalar and pseudoscalar couplings $C_R^b$ 
(see \Sec{NHB:four-quark:ops}), which renormalize both
electromagnetic and  gluonic operators,  the two solutions $A_7(\mu_b)
\sim \pm A_7^\sm(\mu_b)$ are allowed. 
This is the case if Eq.~(\ref{eq:CsCpnoRH}) is satisfied.
%
%
%
We update the NLO analyses of  \cite{Ali:2002jg, Kagan:1998ym}
with the inclusive $b\to s \gamma$ measurement in
\eq{exp:bound:bsgamma} and 
$\branch(B\to X_ce\bar \n_e) = 10.80\%$ and obtain the ranges 
($\mu_b=2.5 \ \GeV$)
\begin{eqnarray}\label{bsg:A7}
-0.36 \leqslant A_7 \leqslant  -0.17 \quad \mbox{or}\quad 
0.21 \leqslant A_7\leqslant  0.42.
\end{eqnarray}
The corresponding correlation between
$A_7$ and $A_8$ can be seen in the left plot of \fig{fig:bsgamma}.
For  $C_R^b(m_W)=0.06$, on the other hand,  the experimental constraints on
$A_7$ are  much weaker (right plot of \fig{fig:bsgamma}).

\subsection{Constraints from \bm $b\to s l^+l^-$}

\begin{figure}
\begin{center}
\includegraphics[scale=0.55]{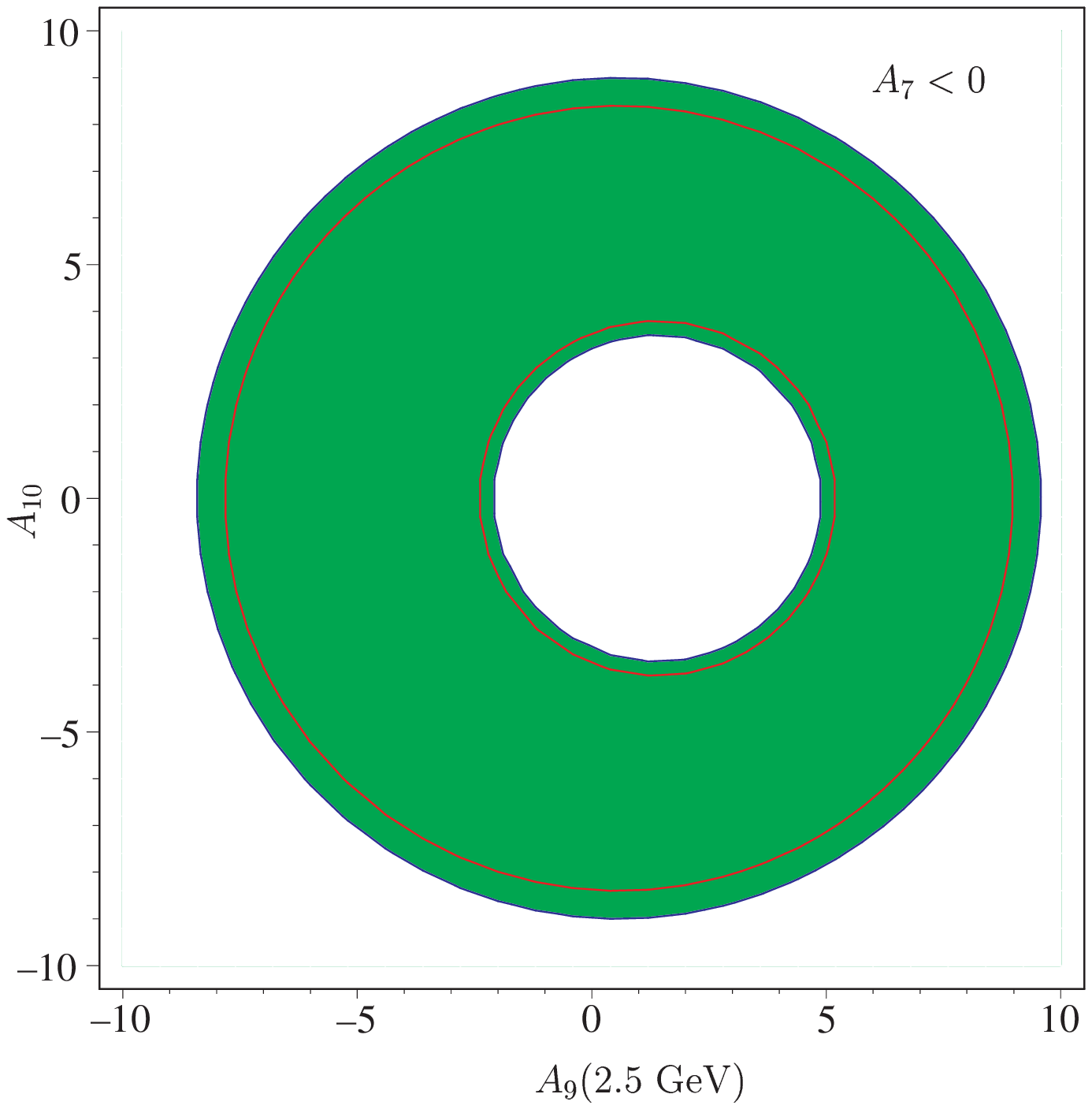}
\hspace{1em}
\includegraphics[scale=0.55]{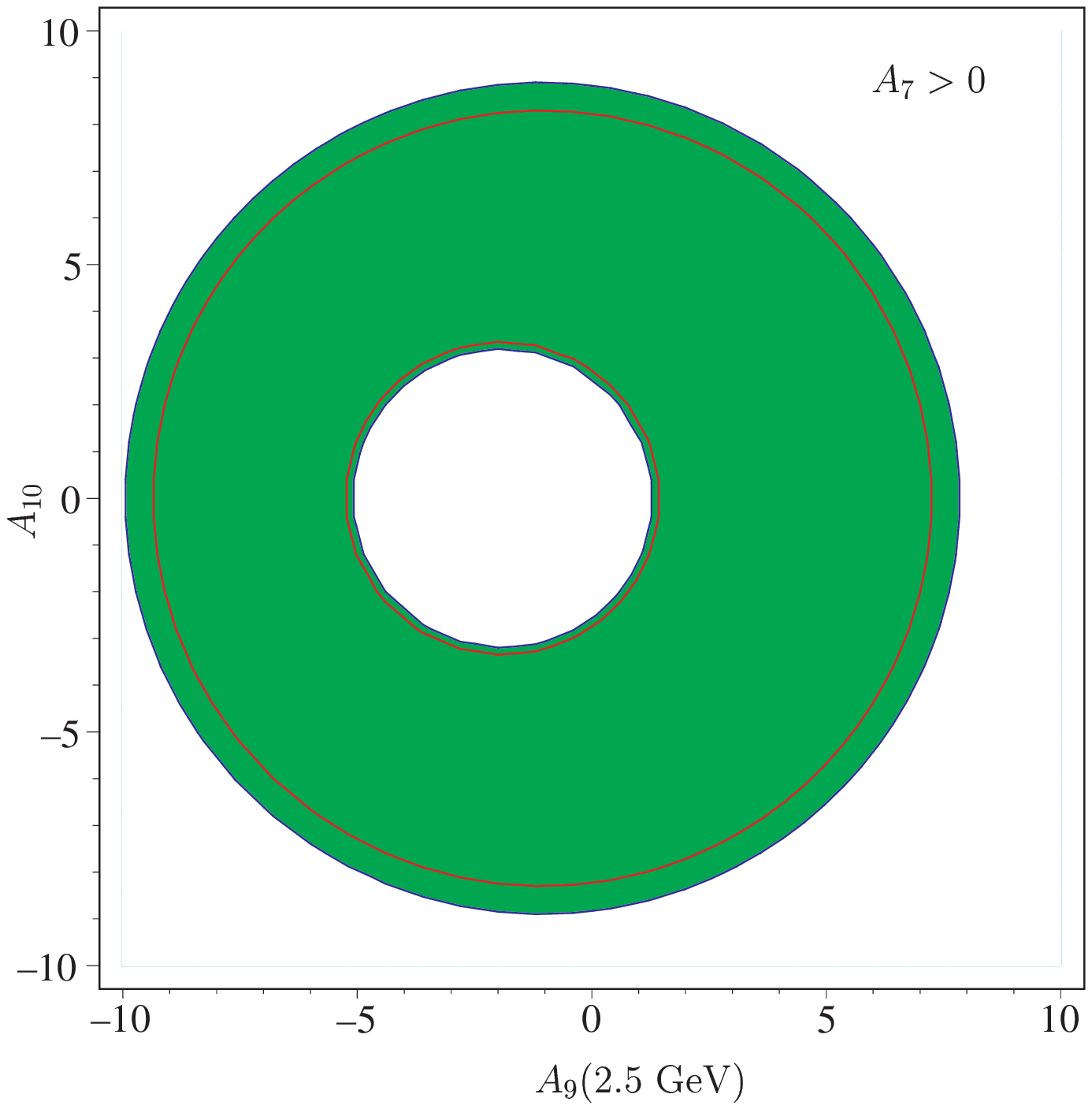}
\vspace{2em}

\includegraphics[scale=0.55]{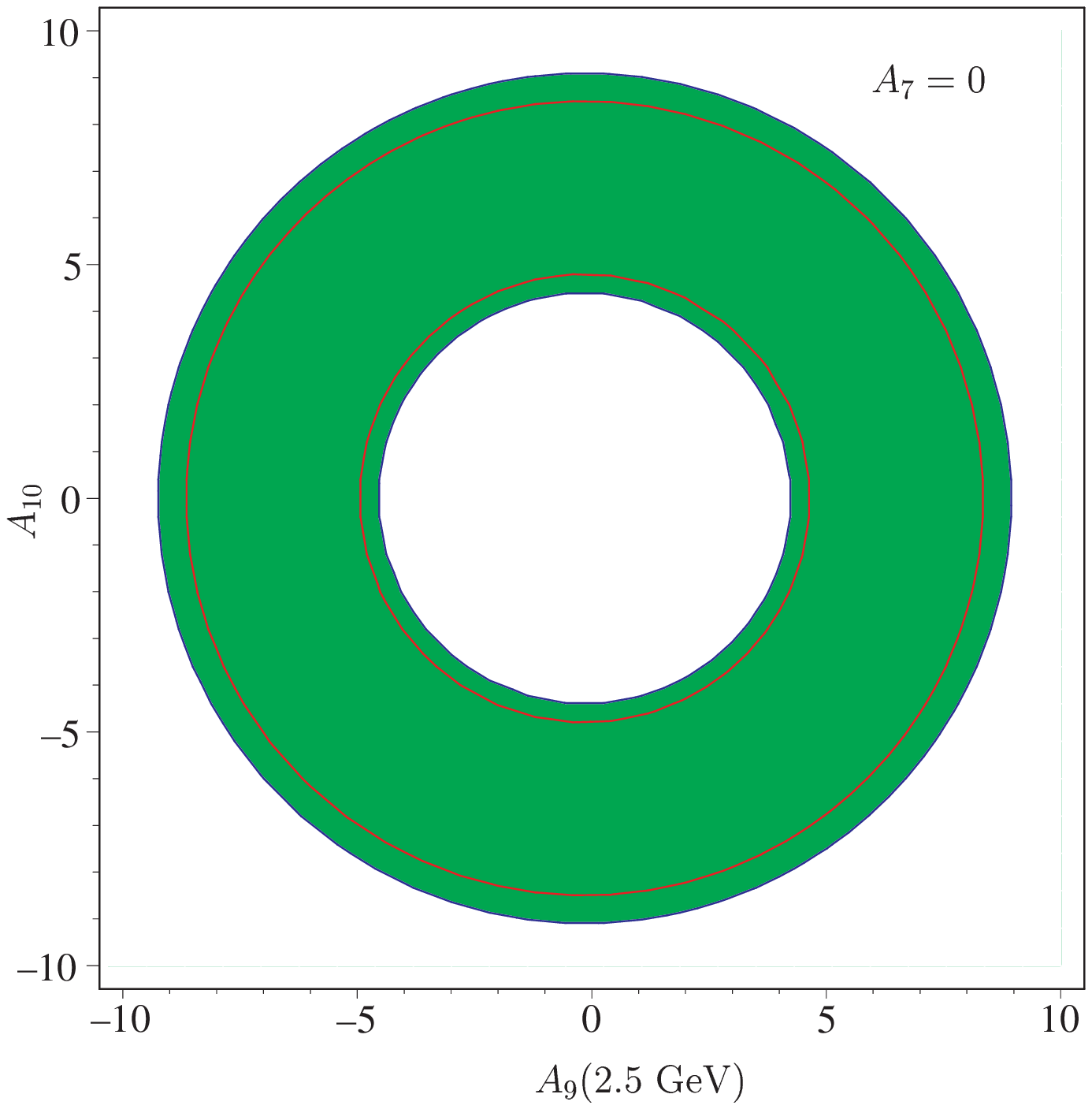}
\end{center}
\caption[]{Allowed regions in the  $A_9$--$A_{10}$ plane in the presence of scalar and pseudoscalar
operators from data on  inclusive $b\to s \ell^+\ell^-$
and $b\to s \gamma$  decays for different values of $A_7$.
The shaded areas are obtained from the upper bound on
$\branch(B\to X_s e^+e^-)$ and
the lower  bound on $\branch(B\to X_s\mu^+\mu^-)$,
 Eqs.~(\ref{eq:up}) and (\ref{lower:bound:mumu})  with
$f_{B_s}=200\ \MeV$.
The two remaining contours  indicate  the allowed  regions from 
 the $\cl{90}$ measurement  of $\branch(B\to X_s e^+e^-)$ given in \eq{eq:up:eeUB}. 
Since the bounds from $\branch(B\to X_s\mu^+\mu^-)$ for
$f_{B_s}=238\ \MeV$ give very similar results, we do not show the
corresponding contours.  
\label{fig:bsee:bsmumu}}
\end{figure}

In the presence of new physics contributions proportional to the
lepton mass we use data on the electron  modes to constrain the dilepton
couplings $A_{9,10}$.
{}From the upper bound on ${\mathcal{B}}(B \to X_s e^+ e^-)$ 
given in \eq{eq:belleXsee} we obtain
\bea\label{eq:up}\nnu
\sqrt{\left|A_9 \mbox{}^{-0.58}_{+1.05}\right|^2
+|A_{10}|^2 } & \leqslant &
\left\{\begin{array}{l}
9.0\quad \mathrm{for }\quad  A_7 <0\\
8.9\quad \mathrm{for }\quad A_7>0
\end{array} \right.\\
\sqrt{\left|A_9 + 0.15  \right|^2 +|A_{10}|^2} & \leqslant &
\hspace{0.95em} 9.1\quad 
\mathrm{for }\quad  A_7 =0.
\eea
The range on  ${\mathcal{B}}(B \to X_s e^+ e^-)$  given in
\eq{eq:Xsee:90CL} yields upper and lower bounds
\bea\label{eq:up:eeUB}
&&\left. \begin{array}{l}
3.8\\
3.3
\end{array}\right\}
\leqslant \sqrt{\left|A_9 \mbox{}^{-0.58}_{+1.05}\right|^2
+|A_{10}|^2} \leqslant
\left\{\begin{array}{l}
8.4\quad \mathrm{for }\quad  A_7 <0\\
8.3\quad \mathrm{for }\quad A_7>0
\end{array}\right.\nnu\\
&& \hspace{0.53cm} 4.8\leqslant \sqrt{\left|A_9 + 0.15  \right|^2 +|A_{10}|^2}  \leqslant  8.5\quad \mathrm{for }\quad  A_7 =0.
\eea
Similar bounds can be obtained from data on the muon modes
together with the upper limit on $C_{S,P}$ in \eq{eq:CSPbound}.
The  lower limit on   ${\mathcal{B}}(B \to X_s \mu^+ \mu^-)$ 
in \eq{eq:Xsmumu:90CL} yields
\bea\label{lower:bound:mumu}
\sqrt{\left|A_9 \mbox{}^{-1.4}_{+1.9} \right|^2 +
|A_{10}|^2}  & \geqslant & 
\left\{\begin{array}{l}
3.8\ (3.5) \quad \mathrm{for}\quad  A_7 <0\\
3.5\ (3.2) \quad \mathrm{for}\quad A_7>0
\end{array} \right. \nnu \\ 
\sqrt{\left|A_9 + 0.15  \right|^2 +
|A_{10}|^2} &  \geqslant & \hspace{0.95em}4.7\ (4.4)\quad \mathrm{for}\quad  A_7 =0
\eea
for  $f_{B_s}=238\ \MeV$ ($200\ \MeV)$.
Our constraints on $A_{9,10}$ given in
\eqs{eq:up}{lower:bound:mumu} are displayed in \fig{fig:bsee:bsmumu}.
Like in the analysis with the restricted SM basis in 
\cite{Ali:2002jg},  $A_9=A_{10} = 0$
is excluded even in
the presence of new scalar and pseudoscalar interactions.

%
%
\subsection{Constraints from \bm$R_K$}\label{subsec:RK}
The experimental bound $R_K \leqslant 1.2 $ in  \eq{ranges:90CL}  
provides
constraints on the scalar and pseudoscalar Wilson coefficients complementary
to those from the $B_s\to
\mu^+\mu^-$ branching fraction given in \eq{eq:CSPbound}. Varying
$A_{7,9,10}$ according to 
Eqs.~(\ref{bsg:A7}), (\ref{eq:up})--(\ref{lower:bound:mumu}) we
obtain   ($\mu_b = 2.5\ \GeV$)
\bea\label{eq:CSPbound:RK}
\sqrt{|C_S|^2+|C_P+\Delta_{10}|^2 } \leqslant
4.5.
\eea
Here, $\Delta_{10}$ stems from the interference term of $C_P$  and $A_{10}$ 
in the $B \to K \mu^+ \mu^-$ rate, see Eq.~(\ref{eq:BKll}),
which can be neglected for  large values of $C_{S,P}$.  
If the bound on $R_K$ improves e.g.~to 1.1, then the value on the r.h.s.~of the above
equation changes to 3.2.

\newsection{Correlation between \bm $B_s\to \mu^+\mu^- $ and $b \to
s\ell^+ \ell^-$ decays}
\label{correlations}
In this section we study correlations between the ratios $R_H$ 
defined in \eq{eq:RH}
and ${\mathcal{B}}(B_s\to \mu^+\mu^-)$. 
We restrict ourselves to the case $C_S=-C_P$, hence a vanishing
$A_7$ is excluded as shown in \Sec{sec:a7a8}.
We further assume that 
$A_{9,10}$ are SM valued
while $A_7$ is allowed to vary in the intervals given in \eq{bsg:A7}. 
This particular scenario is, for example, 
realized in the MSSM with MFV at large $\tan \beta$.
The maximum values of  $R_H$ are summarized in Table
\ref{table:final:results} of \Sec{summary} for different new physics scenarios.

The correlations depend sensitively on  the decay constant of the 
$B_s$ meson. We display our results for $f_{B_s}=200\  \MeV$ and $238\
\MeV$ except for
the inclusive decays, where we vary between these two values.
As described in \Sec{sec:constraints} 
we use the partial NNLO expressions.
Therefore, the plots are obtained for fixed renormalization scale $\mu_b=2.5$ GeV.
For the analysis of the exclusive decays we show 
the uncertainty from the form factors.

The SM predictions for  the ratios $R_H$  
are summarized in Eqs.~(\ref{eq:sm1}) and (\ref{eq:sm2}).
The theoretical uncertainty for the inclusive decays 
is due to the variation of the
renormalization scale between $2.5\ \GeV$  and $10\ \GeV$. 
Since we are using the partial NNLO expressions
this small error below one  percent on $R_{X_s}^{SM}$ might even be
overestimated. For comparison, we give the corresponding numbers at NLO
$R_{X_s}^{\sm,  \mathrm{NLO}} = 0.974\pm 0.006$  and 
$ R_{X_s}^{\sm, \mathrm{NLO}}|_{\mathrm{low\ } q^2} = 0.972\pm 0.005$.
The SM prediction for the $B_s \to \mu^+ \mu^-$ branching ratio is  $(3.6 \pm 1.4) \times 10^{-9}$,
where the main theoretical uncertainty results from the $B_s$ decay
constant. It can be considerably  reduced  
once the $B_s^0$--$\bar{B}^0_s$ mass difference is known \cite{Buras:Bsmumu}.

\subsection{Exclusive \bm $B \to K  \ell^+ \ell^-$ decays}\label{sec:BK}
\begin{figure}
\begin{center}
\includegraphics[scale=0.55]{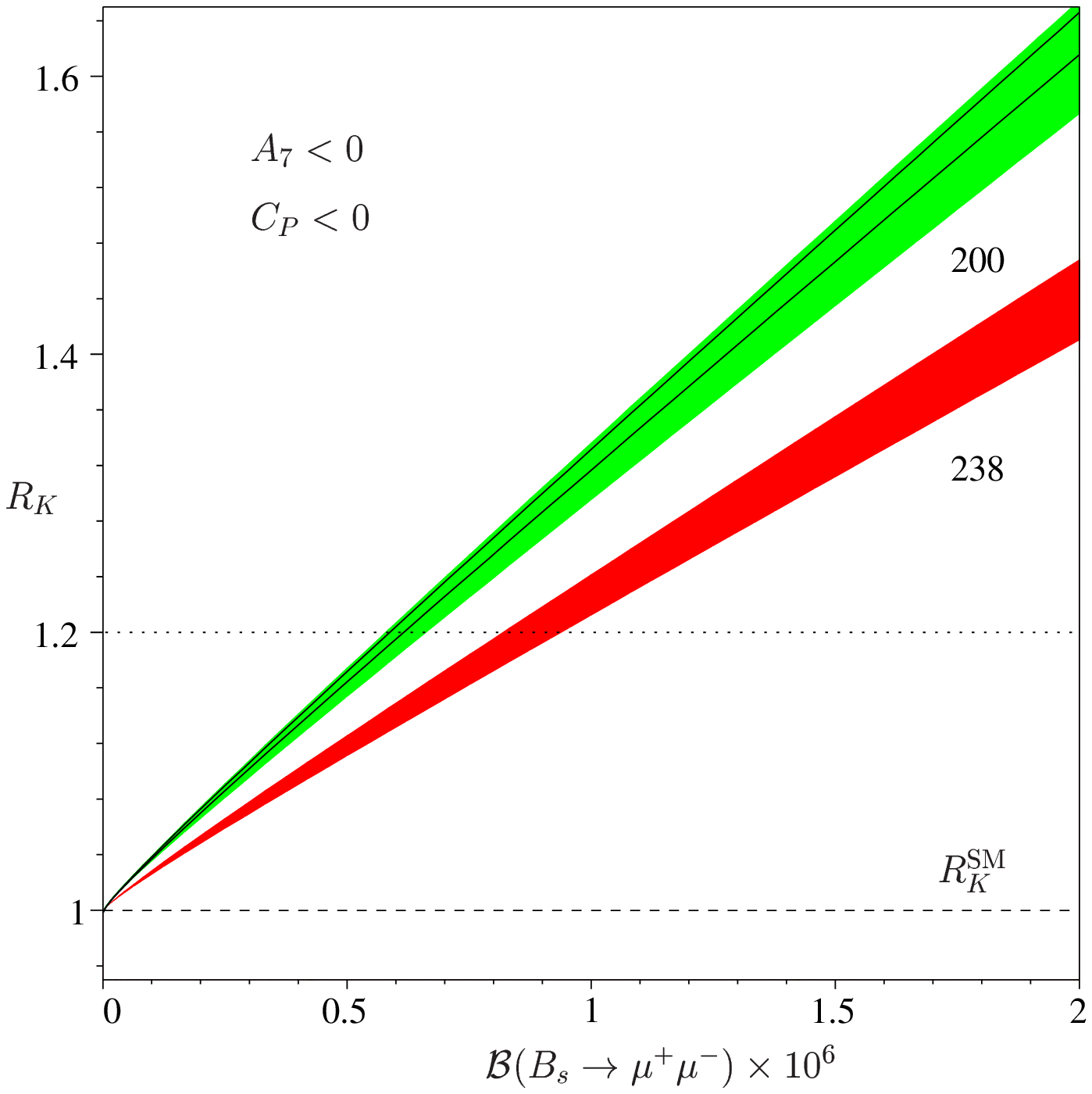}
\hspace{0.8em}
\includegraphics[scale=0.55]{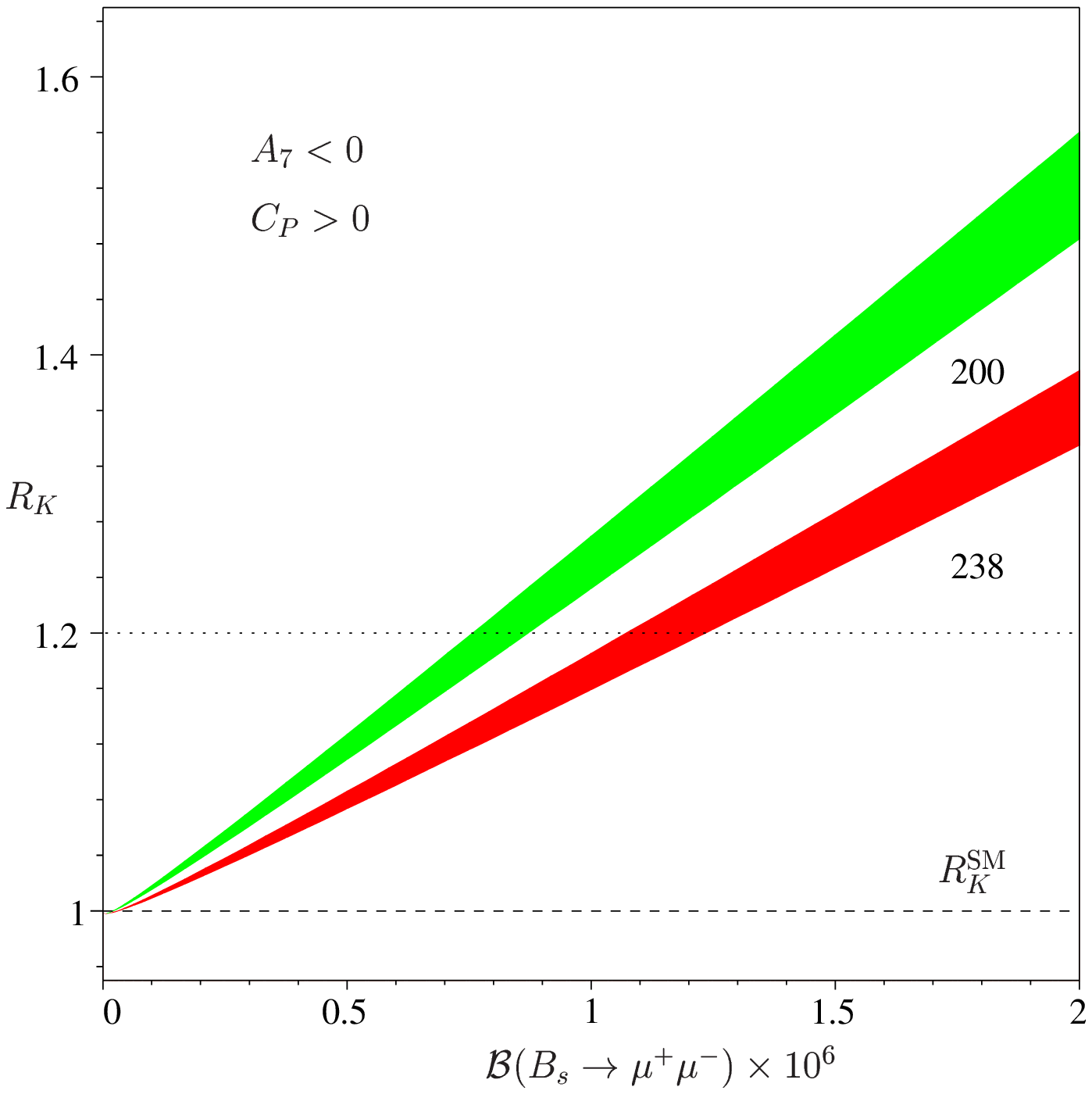}
\vspace{1.5em}

\includegraphics[scale=0.55]{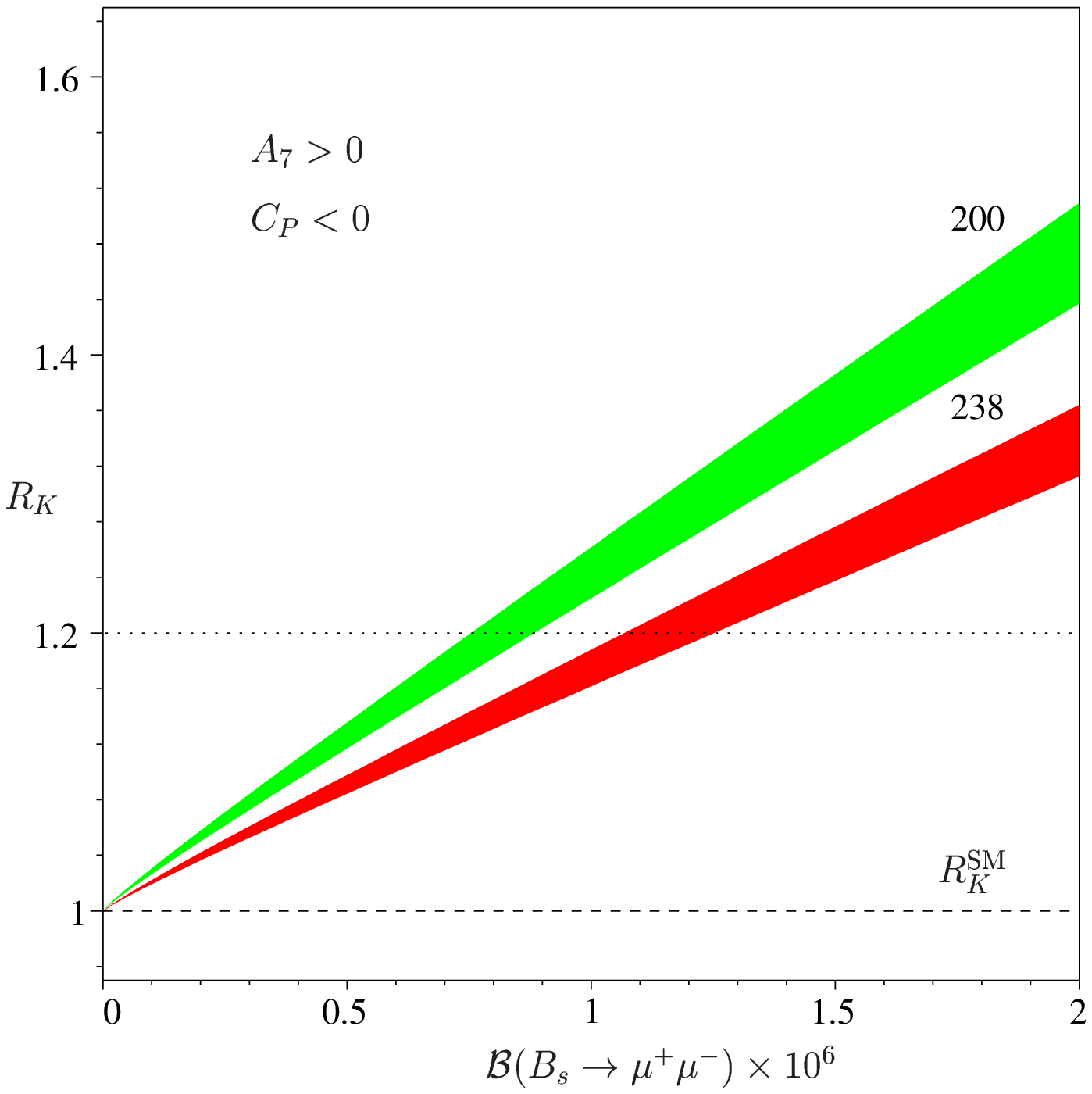}
\hspace{0.8em}
\includegraphics[scale=0.55]{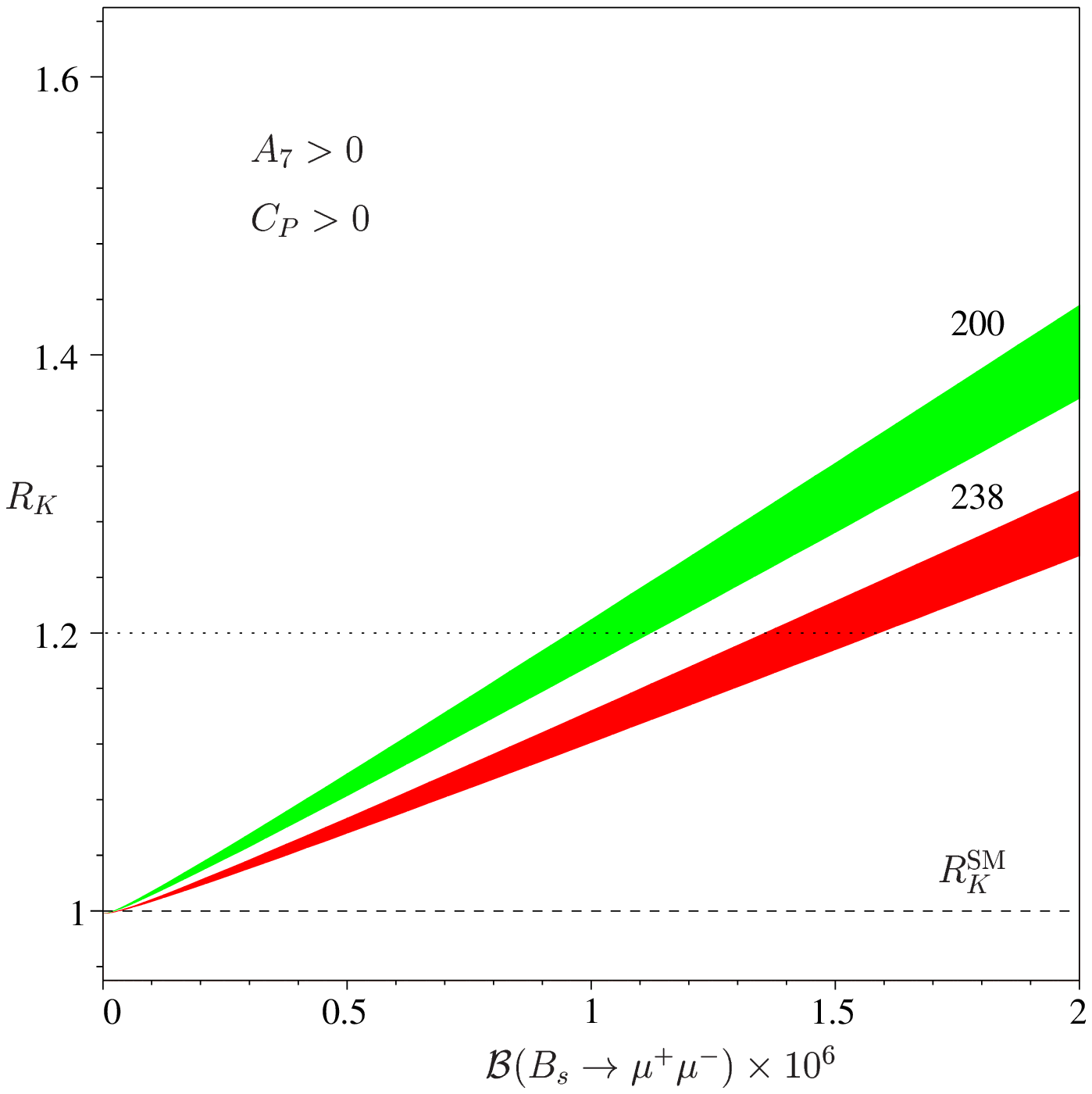}
\end{center}
\vspace{-1.5em}

\caption[]{Correlation between  $R_{K}$ and the $B_s \to \mu^+\mu^-$
branching ratio for different signs  of $A_7$ and
$C_P$, two values of $f_{B_s}$ in $\MeV$ and  $A_{9,10}=A_{9,10}^{\sm}$.
The shaded areas have been obtained by varying the $B\to K$ form
factors according to  \rf{Ali:1999mm} and $A_7$ as given in \eq{bsg:A7}.
In the upper left plot, the form factor uncertainty is illustrated for
fixed $A_7=A_7^\sm$ and $f_{B_s}=200\ \MeV$ by solid lines.~The dotted lines correspond
to the $\cl{90}$ upper limit  on  $R_K$ in \eq{ranges:90CL}.
Dashed lines denote the SM prediction for  $R_K$.\label{fig:RBK}}
\end{figure}

Figure \ref{fig:RBK} shows the correlation between $R_{K}$ and 
the $B_s\to \mu^+\mu^-$ branching ratio  for two values of the
$B_s$-meson decay constant and different signs of $A_7$ and $C_P$.
As illustrated by the solid lines in the upper left plot, the dependence of $R_K$ on the form factors is very small
and hence this observable is useful 
for testing the SM. For comparison, the 
uncertainty on the $B\to K\ell^+\ell^-$
branching fraction  due to the  form factors is $\sim 30 \%$ \cite{Ali:2002jg}.
While being consistent with $B_s\to
\mu^+\mu^-$ data  given in \eq{eq:cdfbound},
an enhancement of $R_K^{\sm}$ by
$\sim 60\%$ is  excluded by the current upper limit on $R_K$ (dotted lines).

%
%
%
%

Furthermore, the ratio $R_K$ provides  a
bound on $C_{S,P}$  which  is competitive with  the limit  from 
$\branch(B_s\to\mu^+\mu^-)$ in \eq{eq:CSPbound}. For two values
of $R_K$  we find   ($\mu_b=2.5\ \GeV$)
\bea\label{eq:CSPbound:RK:A910SM}
\sqrt{|C_S|^2+|C_P-0.4|^2 } \leqslant
\left\{\begin{array}{l}3.2  \quad \mathrm{for}\quad  R_K=1.2\\
2.3 \quad \mathrm{for}\quad R_K=1.1,
\end{array}\right.
\eea
whereas data on $B_s\to\mu^+\mu^-$ decays give
\bea\label{eq:CSPbound:A910SM}
\sqrt{|C_S|^2+|C_P-0.15|^2 } \leqslant 3.8 \Bigg[\frac{{\mathcal{B}}( B_s \to \mu^+ \mu^-)}{2.0 \times
10^{-6}}\Bigg]^{1/2}\Bigg[\frac{238\ \MeV}{f_{B_s}}\Bigg]. 
\eea
We recall that $R_K=1.2$ corresponds to the current 
$\cl{90}$ upper limit, see \eq{ranges:90CL}.

\subsection{Exclusive \bm $B \to K^* \ell^+ \ell^-$ decays}\label{sec:BKstar}

\begin{figure}
\begin{center}
\includegraphics[scale=0.55]{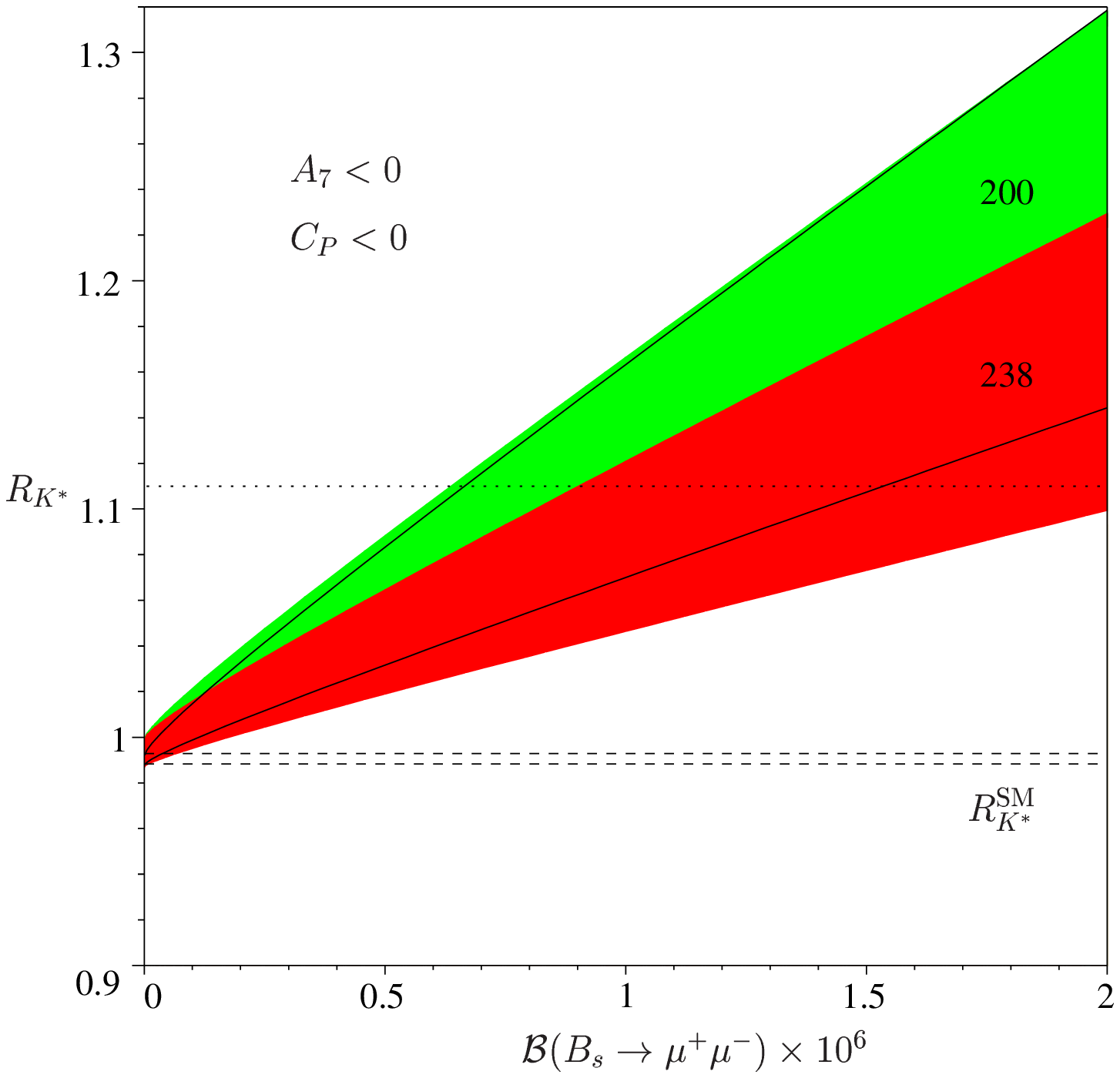}
\hspace{0.8em}
\includegraphics[scale=0.55]{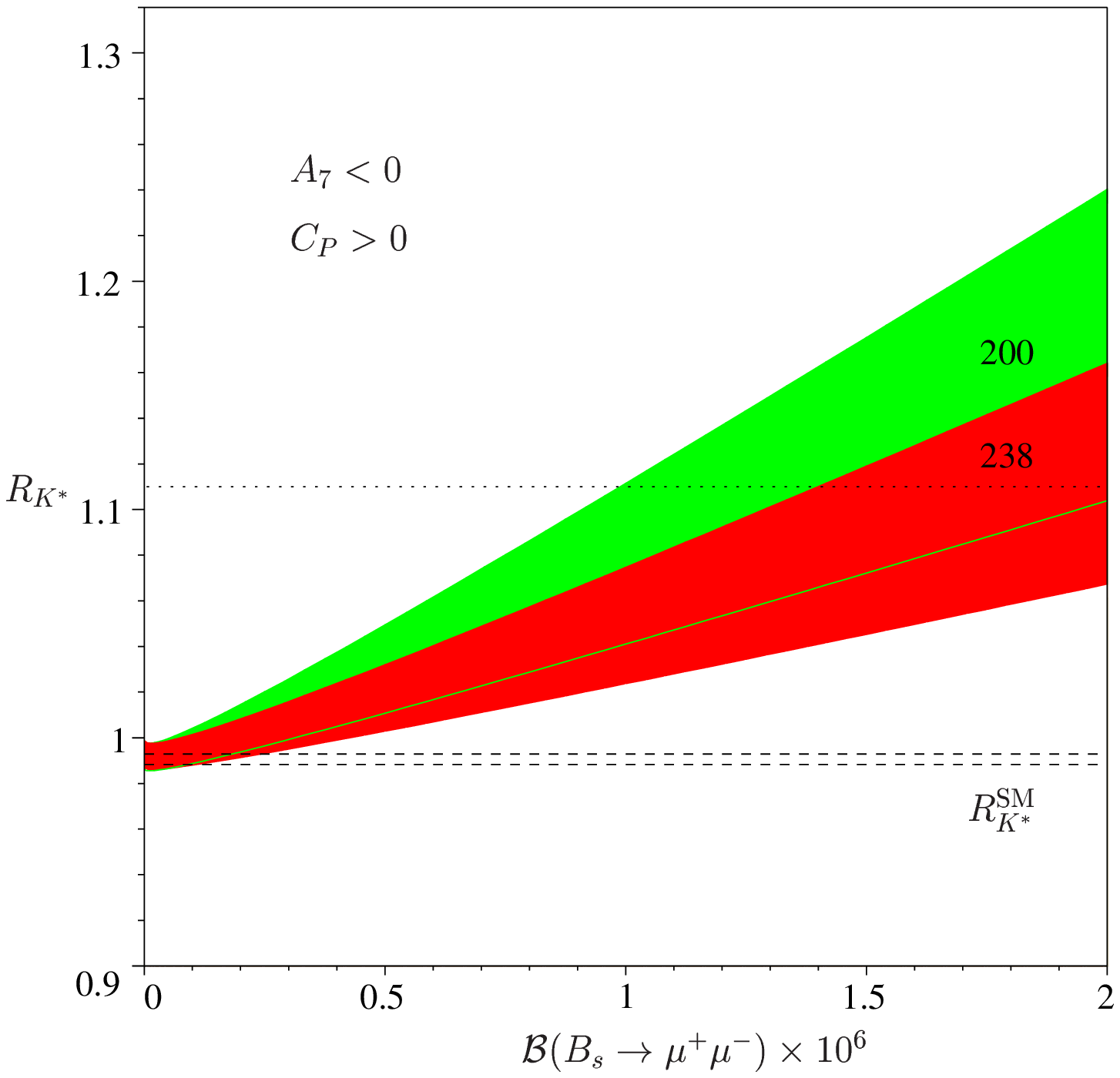}
\vspace{1.5em}

\includegraphics[scale=0.55]{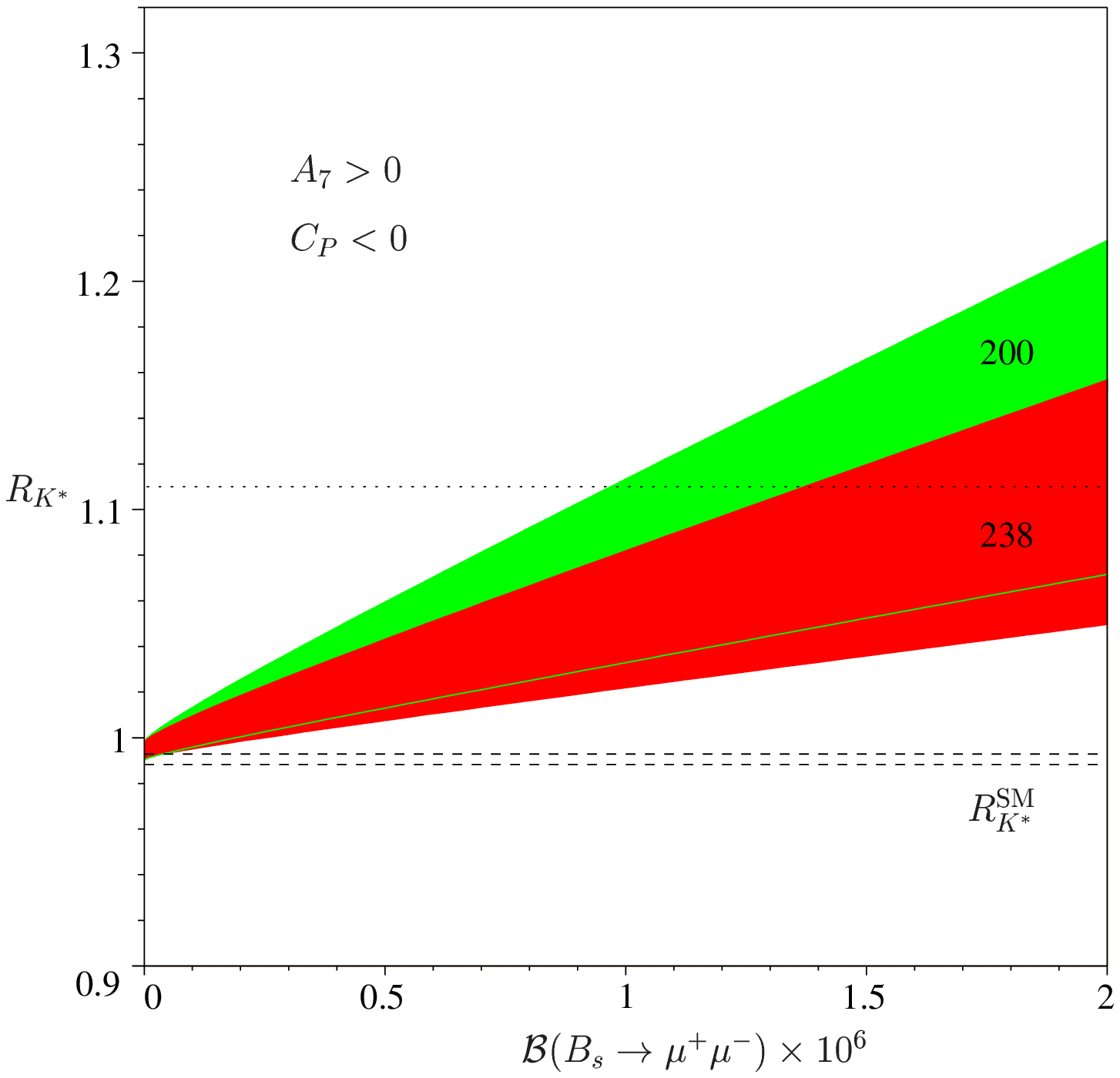}
\hspace{0.8em}
\includegraphics[scale=0.55]{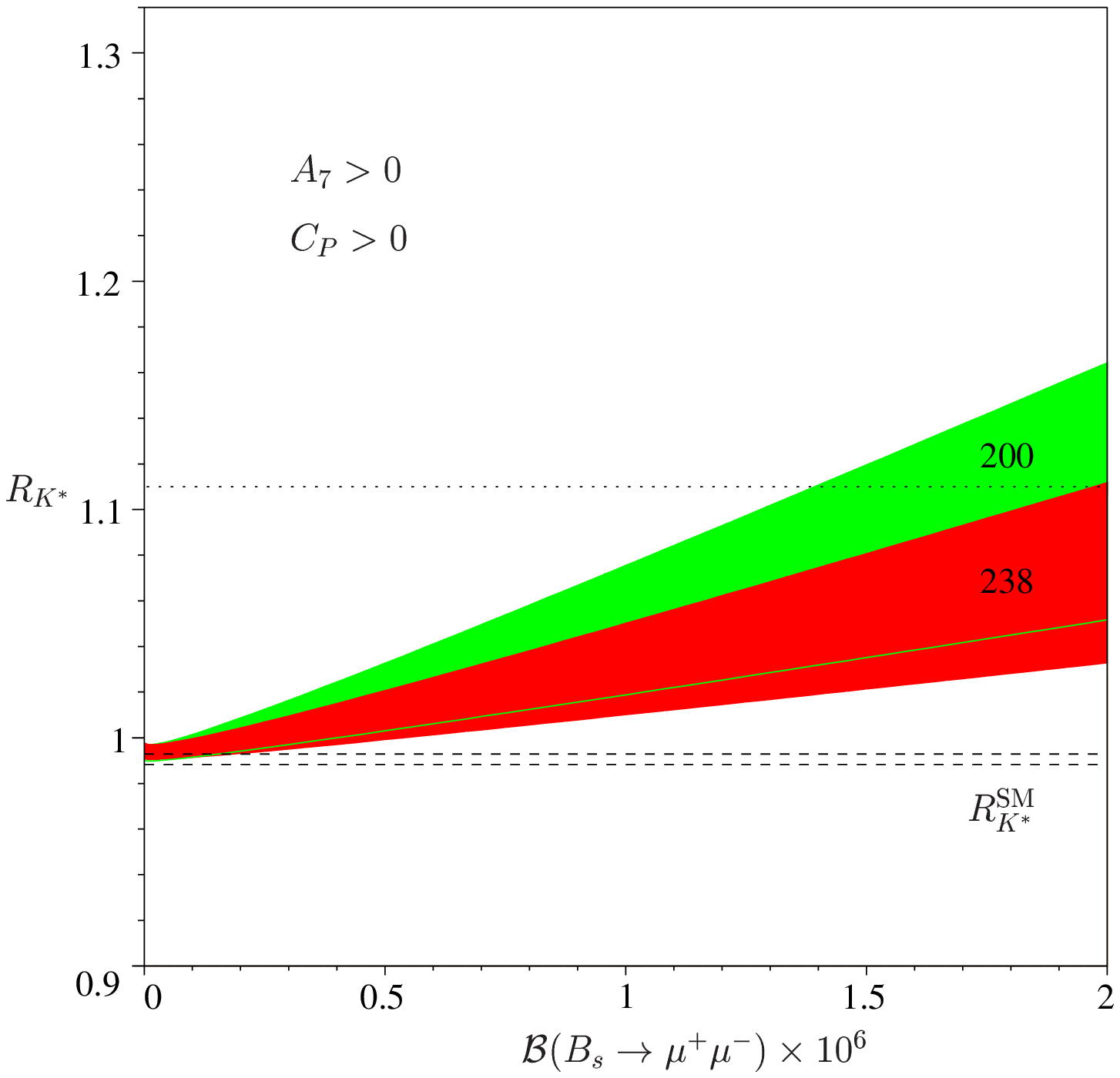}
\end{center}
\vspace{-1.5em}

\caption[]{Correlation between  $R_{K^*}$ and the $B_s \to \mu^+
\mu^-$ branching ratio (see \fig{fig:RBK} for details).
The dotted lines represent the maximum value of $R_{K^*}$ consistent
with the experimental upper limit $R_{K}\leqslant 1.2$.
Dashed lines denote the SM prediction for  $R_{K^*}$.
\label{fig:RBKstar}}
\end{figure}

The results for $R_{K^*}$ versus the branching ratio  of $B_s\to \mu^+\mu^-$ 
are shown in \fig{fig:RBKstar}. 
Note that the variation from the form factors is much larger
than in $R_K$.
This is caused by the form factor $A_0$, which drives the $C_{S,P}$ 
contributions to $R_{K^*}$. Its theoretical uncertainty in light cone QCD 
sum rules \cite{Ali:1999mm}, which we use in our analysis, is twice 
as large as in $f_0$ relevant for $R_K$. 
New physics effects in  $R_{K^*} $
can be as large as  $30\%$ 
[allowed by $B _s \to \mu^+ \mu^-$ data in \eq{eq:cdfbound}] but are 
restricted to be less than $12\% $
once data on $R_{K}$ are taken into account.
For the ratio with
no lower cut on the electron mode we find including all constraints
$R_{K^*}|_{\mathrm{no\ cut}} \leqslant  1.01$, an enhancement of $36 \%$ over its
SM value.

%
%
%
%

\subsection{Inclusive \bm $B \to X_s \ell^+ \ell^-$ decays \label{sec:BXs}}

\begin{figure}
\begin{center}
\includegraphics[scale=0.55]{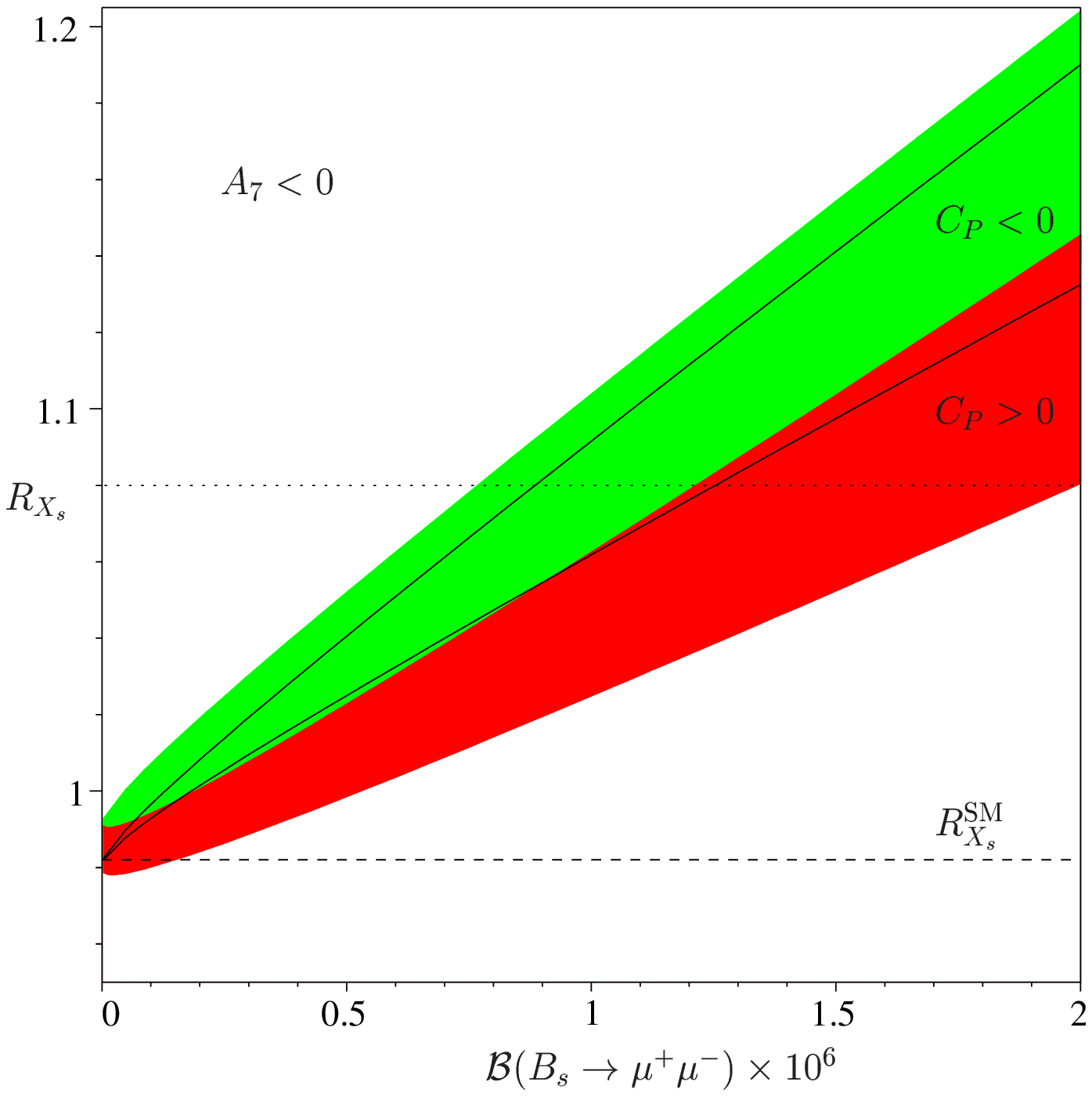}
\hspace{1em}
\includegraphics[scale=0.55]{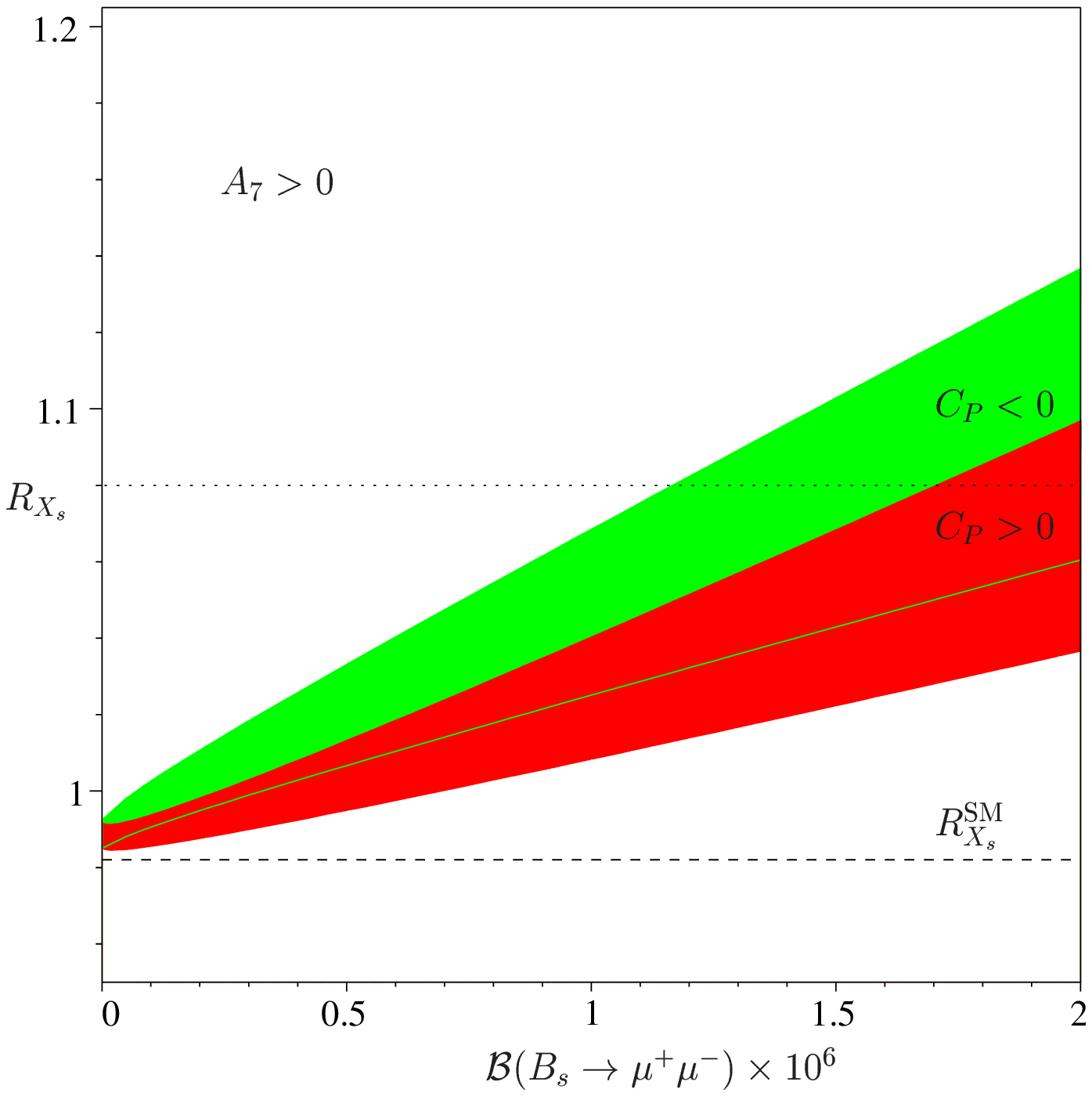}
\vspace{1.5em}

\includegraphics[scale=0.55]{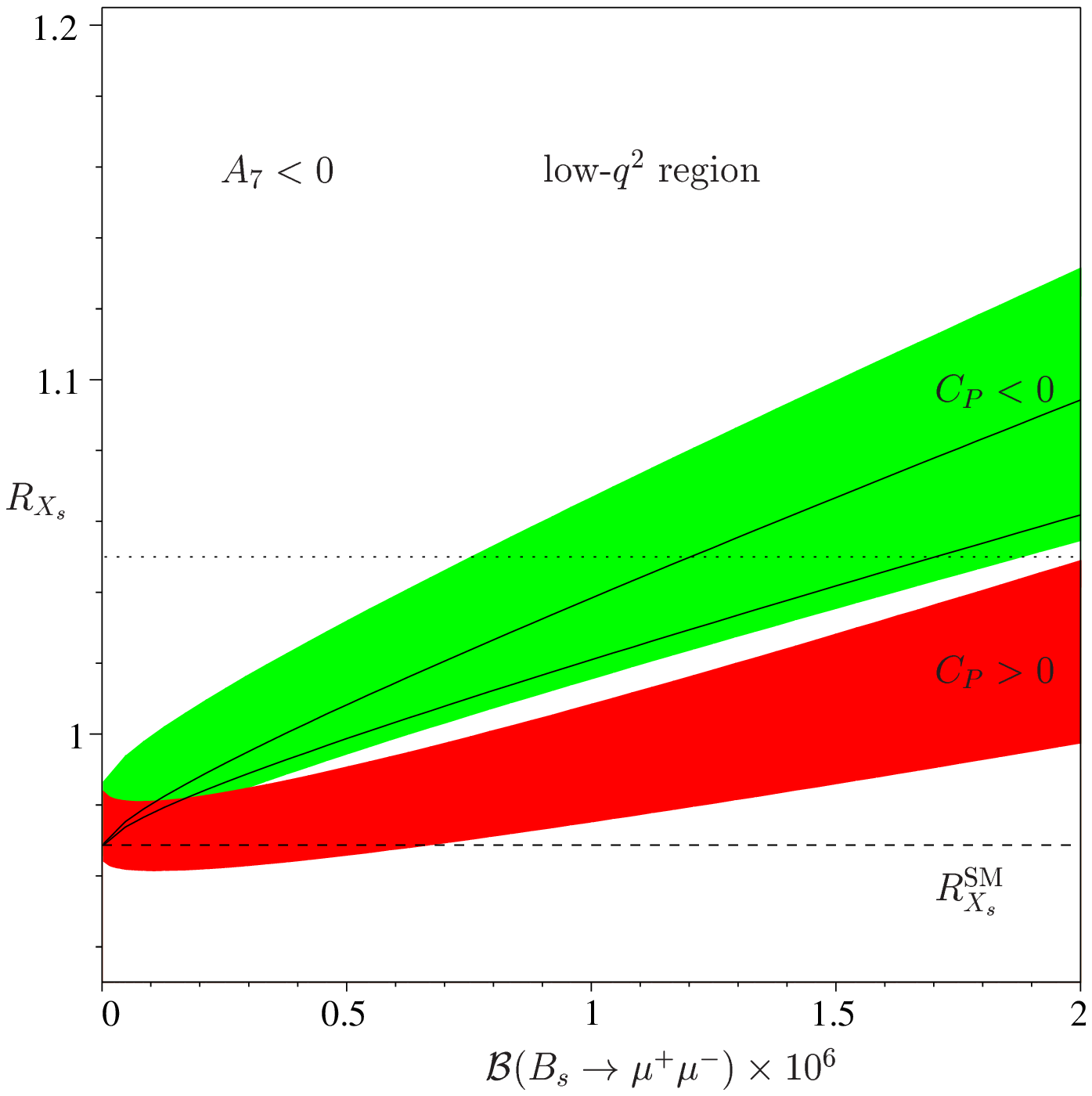}
\hspace{1em}
\includegraphics[scale=0.55]{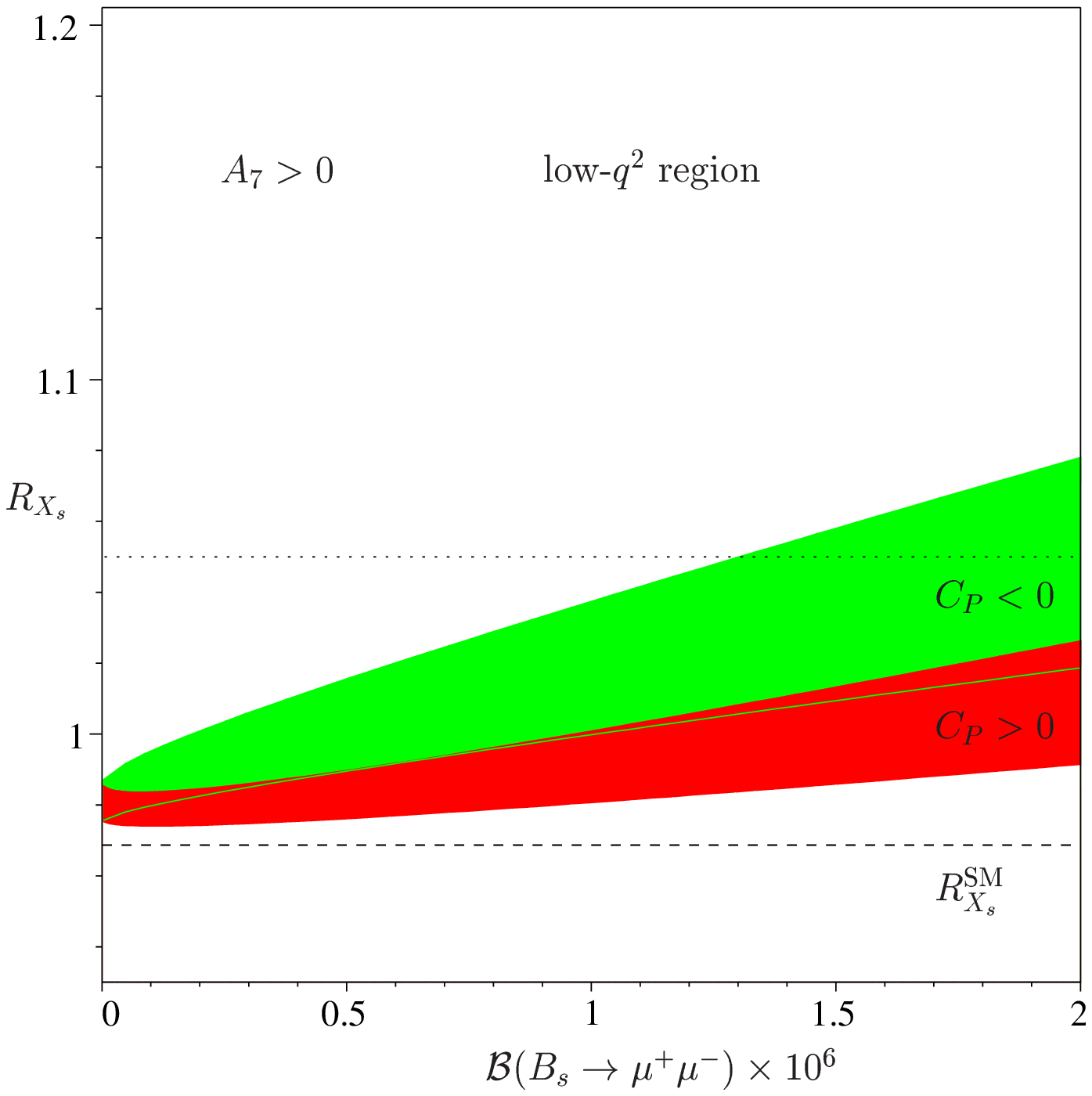}
\end{center}
\vspace{-1.5em}

\caption[]{The dependence  of $R_{X_s}$ on the 
$B_s \to \mu^+ \mu^-$ branching ratio for  different signs of
$A_7$ and $C_P$, and  $A_{9,10}=A_{9,10}^{\sm}$. The upper plots
    correspond to  the whole dilepton invariant mass  spectrum while  the lower
    ones correspond to the  low-$q^2$  region as described in the text.
The shaded areas have been obtained by varying $f_{B_s}$ between $200\ \MeV$ and
$238\ \MeV$ and $A_7$ according to \eq{bsg:A7}. In the
left  plots, the solid lines indicate the uncertainty from the
variation of $f_{B_s}$ for fixed  $A_7 = A_7^\sm$ and $C_P<0$.  The
dotted lines represent
the maximum allowed values of
$R_{X_s}$ obtained for $R_K\leqslant 1.2$.
Dashed lines denote the SM prediction for  $R_{X_s}$.
\label{fig:RBX}}
\end{figure}

In \fig{fig:RBX} we show the correlation of $R_{X_s}$ with
the $B_s\to \mu^+\mu^-$ branching ratio
for the full spectrum with $\hat s_{\max}\approx 1$ (upper plots) and 
for the low dilepton mass  with $\hat s_{\max}=0.26$ (lower plots). 
Order one effects in $R_{X_s}$ from scalar and pseudoscalar interactions 
are excluded
by current data on $B_s\to \mu^+\mu^-$, contrary to the results of
\rf{Wang:2003je} but  in agreement with \rf{piotr:03}.
We find a maximum value of  $R_{X_s}$ of $1.08$ (full spectrum) and 
$1.05$ (low dilepton mass) from the experimental upper limit on $R_K$.  
These bounds on the 
$B\to X_s\mu^+\mu^-$ branching ratio are similar to the ones 
from $B_s\to \mu^+\mu^-$ data previously obtained in 
\cite{piotr:03}. 
While an enhancement of the $B\to X_s\mu^+\mu^-$ branching ratio 
of $O(10\%)$ is within the uncertainty of the SM prediction,
a  corresponding  effect in the ratios $R_{X_s}$ can be well
distinguished from the SM ones.

%
%
%

\newsection{Summary}\label{summary}
We performed for the first time a  
model-independent analysis of $b\to s$ processes
in an extended operator 
basis, the SM one  with ${\mathcal{O}}_{7-10}$ plus scalar and
pseudoscalar operators ${\mathcal{O}}_{S,P}$ with dileptons.
In our phenomenological analysis 
we took into account experimental constraints from inclusive
$b \to s \gamma$, $b \to s \ell^+ \ell^-$ ($\ell=e,\mu$) and 
$B_s \to \mu^+ \mu^-$ decays. 
Further, we used data on the ratio of $B \to K \mu^+
      \mu^-$ to $B \to K e^+ e^-$  branching ratios, $R_K$.
We made a few assumptions to facilitate this analysis:  no right-handed 
currents, the couplings to the scalar and pseudoscalar 
operators are driven by the respective
fermion mass   and no  CP violation beyond 
the CKM matrix.

We studied the effects of scalar and pseudoscalar operators involving 
$b$ quarks ${\mathcal{O}}^b_{L,R}$. 
Already at zeroth order in the strong coupling constant these operators 
mix onto the SM basis:
${\mathcal{O}}^b_{L}$ proportional to $C_S-C_P$ onto the 
4-Fermi operators with dileptons 
and 
${\mathcal{O}}^b_{R}$ proportional to $C_S+C_P$ onto the photonic and gluonic 
dipole operators.
Furthermore, we find that the QCD penguins get renormalized at $O(\alpha_s)$ by 
${\mathcal{O}}^b_{L}$. While being negligible in  $b \to s \ell^+
\ell^-$,
these corrections are important for hadronic $b$ decays. In
particular, they cancel the strong $\mu$ dependence of the $B\to \phi
K_S$ amplitude reported recently in \rf{Cheng:2003im}.
The lowest order anomalous dimensions involving ${\mathcal{O}}^b_{R}$
and the dipole operators have been calculated before 
in Ref.~\cite{Borzumati:1999qt},  whereas the ones with ${\mathcal{O}}^b_{L}$
and the 4-Fermi operators
are a new result of this work.
Numerically, we find that for
$C_S=-C_P$ the effects 
of ${\mathcal{O}}^b_{L,R}$ are negligible for our model-independent analysis.
However, for $C_S \neq -C_P$ there is a significant 
impact from  scalar and pseudoscalar couplings 
on the dipole operators. In particular, the branching ratio for the 
decay $b \to s \gamma$ can be obtained completely without any contribution
from the electromagnetic dipole operator 
${\mathcal{O}}_7$. This rather extreme scenario could be excluded 
by improved 
data on the $b \to s g$ branching ratio, as illustrated in 
Fig.~\ref{fig:bsgamma}.
Except for the case of $A_7 \simeq 0$, the bounds we obtain on 
the coefficients $A_{9,10}$ are similar to previous results  in  the 
SM operator basis \cite{Ali:2002jg}.
The non-trivial renormalization effects we encountered 
show that a model-independent analysis can be quite involved in an 
enlarged operator basis.
%

%
\begin{table}
\begin{center}
\caption{Upper bounds on the ratios $R_H$  for different  $C_{S,P}$ scenarios
for $A_{9,10}$ being SM-like and in parentheses without this constraint.
Data on $b \to s \gamma$,  $b \to s \ell^+ \ell^-$,
$B_s \to \mu^+ \mu^-$ and $R_K$ are taken into account.
\label{table:final:results} }
\smallskip

\begin{tabular}{lcccc}\hline\hline
Ratio  & SM& $C_{S,P}=0$ & {$C_S= -C_P$}  &{$C_S\neq -C_P$} \\
\hline \vspace{-1.3em}\\
$R_{K}$  &1.00  &  1.00 (1.00)   & $1.20$ $(1.20)$    
&   $1.20$ $(1.20)$\\
$R_{K^*}$  &  $0.99$ & 1.00 (1.00) &1.11 (1.12)  & 1.12 (1.12)   \\
$\left.R_{K^*}\right|_{\mathrm{no\ cut}}$ 
&  $0.74$& 0.91 (0.97) & 1.01 (1.07)        & 1.11
(1.12)   \\
$R_{X_s}$  & $0.98$ & 0.99 (0.99)& 1.08 (1.08)    & 1.08  (1.08)\\
$\left. R_{X_s}\right|_{\mathrm{low\ } q^2}$       & $ 0.97$ &0.99 (0.99)
 & 1.05 (1.06)           & 1.07 (1.07)\\
\vspace{-1.3em}\\
\hline\hline\end{tabular}
\end{center}
\end{table}
%
%

We worked out correlations between the ratios $R_H$ defined in 
\eq{eq:RH} and the
$B_s \to \mu^+ \mu^-$ branching ratio for
$C_S=-C_P$ and $A_{9,10}$ being SM-like, summarized  in 
\figs{fig:RBK}{fig:RBX}.
This particular scenario  also applies  to the MSSM with MFV at large $\tan \beta$.
Figure \ref{fig:RBK} shows that a bound on $R_K$ implies a bound on
${\mathcal{B}}(B_s \to \mu^+ \mu^-)$ and vice versa.  Current data
on these observables yield very similar constraints on $C_{S,P}$
given in sections \ref{sec:constraints} and \ref{correlations}.
Note that in the above-mentioned MSSM scenario 
${\mathcal{B}}(B_s \to \mu^+ \mu^-)$ and
$B^0_s$--$ \bar{B}^0_s$ mixing are correlated \cite{Buras:2002wq}. 
A similar correlation between $R_K$ and in general with larger
theoretical  errors
also with the other  $R$'s and $B_s^0$--$ \bar{B}^0_s$ mixing then holds
in this model, too.
We stress that in our analysis we take into account
information on branching ratios only from inclusive decays. The data
on exclusive decays enter our analysis only via $R_K$ which
depends only weakly on the form factors, as can be seen from
\fig{fig:RBK}.  The largest theoretical
uncertainty in the correlations  is due to the $B_s$-meson decay constant.

We further calculated the maximal allowed values of the ratios $R_H$, 
summarized in Table \ref{table:final:results}. Since we use the partial NNLO
expressions for the Wilson coefficients, they have been obtained at the scale 
$\mu_b=2.5$ GeV.
We see that large, order one corrections to the
respective SM values are already excluded.~Note that these upper bounds are 
insensitive to $f_{B_s}$ because current data on  $R_K \leqslant  1.2$ are here
more constraining than ${\mathcal{B}}(B_s \to \mu^+ \mu^-)$.
The effect from $C_{S,P}$ on $B \to K$ decays is always bigger than 
on $B \to K^*$ and $B\to X_s$ decays.
The reason is that besides different hadronic matrix elements 
in these decays the photon pole $|A_7|^2/\hat{s}$,
which is absent in the $B \to K$ decay, dominates the rate for very low 
dilepton mass.
The inclusive decay  with the spectrum integrated only over the low
dilepton invariant mass is even less sensitive, since the 
lepton-mass-dependent  contributions are suppressed by small 
$\hat{s}$, see \eq{diff:decay:rate:incl}.

Contributions from scalar and pseudoscalar operators with $V+A$ handedness 
can be included in the $B_s \to \ell^+ \ell^-$ 
branching ratio by $C_{S,P} \to C_{S,P}-C_{S,P}^\prime $
and into the $B \to K \ell^+ \ell^-$ spectrum by
$C_{S,P} \to C_{S,P}+C_{S,P}^\prime $.
Hence, the correlations we presented between 
${\mathcal{B}}(B_s \to \mu^+ \mu^-)$ and $R_K$
break down if both chirality 
contributions $C_{S,P}$ and $C_{S,P}^\prime$ are non-vanishing.
Since $R_K$ constrains the sum and $B_s \to \mu^+ \mu^-$ the difference
of the coefficients, combining these two [Eqs.~(\ref{eq:CSPbound}) and (\ref{eq:CSPbound:RK})] 
yields an upper bound on the magnitude of the individual coefficients of
$|C_{S,P}^{(\prime)}| \leqslant 4.3 $.
This excludes large cancellations and holds even with right-handed contributions
to the SM operator basis.

In conclusion, $b \to s \ell^+ \ell^-$ induced decays can have a splitting
in the branching ratios depending on the final lepton flavor 
from physics beyond the SM. 
Hence, averaging of electron and muon data 
has to be done carefully in order not to yield a model-dependent result.
The effect from scalar and pseudoscalar couplings is best isolated in the 
theoretically clean observables $R_H$ with the
{\it same cuts on the dilepton mass}. On the other hand,
the ratio $R_{K^*}|_{\mathrm{no\ cut}}$  constructed with physical phase space
boundaries is also sensitive to new physics  not residing in
$C_{S,P}$, as can be seen
from Table \ref{table:final:results}.

\emph{Note added.} The lowest order mixing of scalar and pseudoscalar operators onto the
SM basis calculated in \Sec{sec:III:sub} has been taken into account in a
revised version of the first paper of
\rf{Cheng:2003im}.

\section*{Acknowledgments}
We would like to thank Martin Beneke, Christoph Bobeth,
Gerhard Buchalla, Andrzej J.~Buras,
Athanasios Dedes, Thorsten Ewerth, Martin Gorbahn, Alex Kagan  and Thomas Rizzo
for useful discussions. We also thank Andrzej J.~Buras for his
comments on the manuscript.~~G.H.~gratefully acknowledges the hospitality of the 
theory group at SLAC,
where parts of this work have been done.~F.K. would like to thank the theory
group at CFIF, Lisbon for hospitality
while part of this work was done.~This research was supported in part by the
Deutsche Forschungsgemeinschaft  under contract Bu.706/1-2. 
\appendix
\newsection{Standard operator basis}\label{tilde:basis}
In this appendix we give the ``standard'' operator basis \cite{BBL}
\bea\label{app:lange_basis}
\widetilde{\Oi}_1= (\bar{s}_{ \alpha} \gamma_\mu P_L c_{ \beta})(\bar{c}_{
\beta} \gamma^\mu P_Lb_{ \alpha}),\quad\quad
\widetilde{\Oi}_2 = (\bar{s}_{ \alpha} \gamma_\mu P_L c_{ \alpha})
(\bar{c}_{ \beta} \gamma^\mu P_L b_{ \beta}),\nnu
\eea
\bea
\widetilde{\Oi}_3 = (\bar{s}_{ \alpha} \gamma_\mu P_L b_{ \alpha})\sum_{q=u,d,s,c,b}
(\bar{q}_{ \beta} \gamma^\mu P_L q_{ \beta}),\quad\quad
\widetilde{\Oi}_4= (\bar{s}_{ \alpha} \gamma_\mu P_L b_{ \beta})
\sum_{q=u,d,s,c,b}(\bar{q}_{ \beta} \gamma^\mu P_L q_{ \alpha}),\nnu
\eea
\bea
\widetilde{\Oi}_5= (\bar{s}_{ \alpha} \gamma_\mu P_L b_{ \alpha})
\sum_{q=u,d,s,c,b}(\bar{q}_{ \beta} \gamma^\mu P_R q_{ \beta}),\quad\quad
\widetilde{\Oi}_6= (\bar{s}_{ \alpha} \gamma_\mu P_L b_{ \beta})
\sum_{q=u,d,s,c,b}(\bar{q}_{ \beta} \gamma^\mu P_R q_{ \alpha}),\nnu
\eea
\bea
\widetilde{\mathcal{O}}_{7}^{{e}}=\frac{3}{2}( \bar{s}_{\alpha } \gamma_\mu
P_L b_{\alpha }) \sum_{q} Q_q (\bar{q}_{\beta} \gamma^\mu P_R q_{\beta }), \quad
\widetilde{\mathcal{O}}_{8}^{{e}}=\frac{3}{2}( \bar{s}_{\alpha }
\gamma_\mu P_L b_{\beta }) \sum_{q} Q_q (\bar{q}_{\beta} \gamma^\mu
P_R q_{\alpha}), \nnu
\eea
\bea
\widetilde{\mathcal{O}}_{9}^{{e}}=\frac{3}{2}( \bar{s}_{\alpha } \gamma_\mu
P_L b_{\alpha }) \sum_{q} Q_q (\bar{q}_{\beta } \gamma^\mu P_L q_{\beta }), \quad
\widetilde{\mathcal{O}}_{10}^{{e}}=\frac{3}{2}( \bar{s}_{\alpha }
\gamma_\mu P_L b_{\beta }) \sum_{q} Q_q (\bar{q}_{\beta} \gamma^\mu
P_L q_{\alpha}), \nnu
\eea
\bea
\widetilde{\mathcal{O}}_{7} = \frac{e}{16\pi^2} m_b
(\bar{s}_\alpha \sigma_{\mu \nu} P_R b_\alpha) F^{\mu \nu}, \quad 
\widetilde{\mathcal{O}}_{8}=\frac{g_s}{16\pi^2} m_b
(\bar{s}_{\alpha} \sigma_{\mu \nu} T^a_{\alpha \beta} 
P_R b_{\beta})G^{a \mu \nu},\nnu
\eea
\bea
\widetilde{\mathcal{O}}_{9} = \frac{e^2}{16\pi^2} 
(\bar{s}_\alpha \gamma_{\mu} P_L b_\alpha)(\bar \ell \gamma^\mu \ell), \quad 
\widetilde{\mathcal{O}}_{10}=\frac{e^2}{16\pi^2}
(\bar{s}_\alpha  \gamma_{\mu} P_L b_\alpha)(  \bar \ell \gamma^\mu \gamma_5\ell).
\eea
Here  $Q_q$ denotes the charge of the $q$ quark in units of $e$, 
$\a$, $\b$ are color 
indices, $a$ labels  the SU(3) generators, $P_{L,R}= (1\mp \g_5)/2$ and
$m_b=  m_b(\mu)$ is the running mass in the $\ol{\mbox{MS}}$ scheme,
\begin{eqnarray}\label{eq:mbrunning}
 m_b (\mu)= m^{\rm pole}_b \Bigg[1-\frac{\alpha_s(m^{\rm pole}_b)}{4
    \pi}\frac{16}3\Bigg] \Bigg[\frac{\alpha_s(\mu)}{\alpha_s(m^{\rm
      pole}_b)}\Bigg]^{\frac{\gamma_m^{(0)}}{2 \beta_0}}
    \Bigg\{1+\Bigg[\frac{\gamma_m^{(1)}}{2 \beta_0}-\frac{\beta_1
        \gamma_m^{(0)}}{2\beta_0^2}\Bigg]
\frac{\alpha_s(\mu)-\alpha_s(m^{\rm pole}_b)}{4 \pi}\Bigg\},\nnu\\
\end{eqnarray}
with  $\gamma^{(0)}_m=8$, $\gamma_m^{(1)}=1012/9$,
$\beta_0=23/3$, $\beta_1 = 116/3$.

\newsection{New operators and mixing}\label{sec:mixing:ops}
The new physics operators containing scalar,
pseudoscalar and tensor interactions are written as
\bea
\widetilde{\Oi}_{11}= (\bar{s}_\alpha P_R b_\alpha)  (\bar{b}_\alpha P_L b_\alpha),\quad\quad
\widetilde{\Oi}_{12}= (\bar{s}_{ \alpha} P_R b_{ \beta} )(\bar{b}_{ \beta}
P_Lb_{ \alpha} ) ,\nnu
\eea
\bea
\widetilde{\Oi}_{13}= (\bar{s}_\alpha P_R b_\alpha)  (\bar{b}_\alpha P_R b_\alpha),\quad\quad
\widetilde{\Oi}_{14}= (\bar{s}_{ \alpha} P_R b_{ \beta} )(\bar{b}_{ \beta} P_R
b_{ \alpha} ) ,\nnu
\eea
\bea\label{add:operators}
\widetilde{\Oi}_{15}= (\bar{s}_\alpha\sigma_{\mu\nu} P_R b_\alpha)  (\bar{b}_\alpha\sigma^{\mu\nu}P_R b_\alpha),\quad\quad
\widetilde{\Oi}_{16}= (\bar{s}_{ \alpha}\sigma_{\mu\nu}  P_R b_{ \beta} )(\bar{b}_{ \beta}\sigma^{\mu\nu}  P_R
b_{ \alpha}),
\eea
where $\sigma_{\mu\nu} = (i/2) [\gamma_\mu,\gamma_\nu]$ and
$\widetilde{\Oi}_{11,13}\equiv  {\Oi}_{L,R}^b$ in \eq{new:ops:scalar}.
For completeness, we give their lowest order self mixing
\cite{Buras:2000if,Borzumati:1999qt,Huang:1999im}, i.e., 
among $\widetilde{\Oi}_{11}, \widetilde{\Oi}_{12}$
\bea\label{ad:11:8}
\gamma = \frac{\alpha_s}{4\pi}\left(\begin{array}{cc}
-16 & 0 \\
-6& 2
\end{array}
\right)
\eea
and  among $\widetilde{\Oi}_{13}, \dots, \widetilde{\Oi}_{16}$
\bea
\gamma =\frac{\alpha_s}{4\pi}
\left(\begin{array}{cccc}
-16  & 0     & {1}/{3}  & -1 \\
 -6   & 2     & -{1}/{2}  & -{7}/{6}\\
 16   & -48  & {16}/{3}  &0 \\
 -24  & -56  & 6  & -{38}/{3}
\end{array}
\right).
\eea
We obtain the following lowest order anomalous dimensions for the
mixing of  $\widetilde{\Oi}_{13}, \dots, \widetilde{\Oi}_{16}$ 
onto  $\widetilde{\Oi}_{7,8}$
\bea
\label{eq:AD}
\gamma_{13-16, 7} = Q_d (1,N_c, - [ 4+8N_c], -[4N_c +8]),
\quad 
\gamma_{13-16, 8} =   (1,0,-4,-8),
\eea
and of $\widetilde{\Oi}_{11,12}$  
onto $\widetilde{{\Oi}}_{3}, \dots, \widetilde{{\Oi}}_{6},\widetilde{\Oi}_{9}$
\bea
\gamma_{11, 3-6} = \frac{\alpha_s}{4\pi}\frac{1}{3}\left(\frac{1}{N_c},-1,\frac{1}{N_c},-1\right), 
\quad \gamma_{12, 3-6 } = 0 ,\quad 
\gamma_{11, 9} = \frac{2Q_d}{3}, \quad  \gamma_{12, 9} =N_c  \frac{2Q_d}{3},
\eea
where  $N_c$ is the number of colors. Note that Eq.~(\ref{eq:AD}) is in agreement with \cite{Borzumati:1999qt}.
For the mixing of $\widetilde{\mathcal{O}}_{11, 12}$ onto the electroweak penguins
$\widetilde{\mathcal{O}}_{7}^{{e}}, \dots,
\widetilde{\mathcal{O}}_{10}^{{e}}$, we find 
\bea
\gamma^e_{11, 7-10} = -\frac{\alpha}{4\pi}\frac{4Q_d}{9}
(1,0,1,0),\quad 
\gamma^e_{12, 7-10} = -  \frac{\alpha}{4\pi}N_c\frac{4Q_d}{9}  
(1,0,1,0).
\eea
The remaining leading order anomalous dimensions vanish.

\newsection{Differential decay distributions}\label{dist}

We neglect the $s$-quark mass and introduce the
notation 
\bea
\hat{m}_i &=&m_i/m_B, \quad \hat{s} =q^2/m_B^2, \quad  
 \hat{u}(\hat{s}) = \sqrt{\lambda \left(1- \frac{4 \hat{m}_\ell^2}{\hat{s}}\right)},\nnu\\
\lambda &\equiv & 1+\hat{m}_{K^{(*)}}^4+\hat{s}^2-2 \hat{s} -2
\hat{m}_{K^{(*)}}^2 (1+\hat{s})
\eea
for the exclusive decays and
\bea
\hat{m}_i &=&m_i/m_b^{\pole}, \quad \hat{s} =q^2/(m_b^{\pole})^2
\eea
 for the inclusive modes. Then, the various decay distributions in the presence of scalar and
pseudoscalar operators can be written as follows.

\subsection{\bm$B_s \to \ell^+ \ell^-$}
\begin{eqnarray}
\label{eq:brbmm}
\Gamma( B_s \to \ell^+ \ell^-)&=& \frac{G_F^2   \aem^2 m_{B_{s}}^3 f_{B_{s}}^2}{64 \pi^3}
|V_{tb}^{}V_{ts}^\ast |^2 \sqrt{1- \frac{4 m_\ell^2}{m_{B_s}^2}} \nnu \\ 
&  \times &  \Bigg\{ 
\Bigg(1-\frac{4 m_\ell^2}{m_{B_s}^2}\Bigg) 
\Bigg| \frac{m_{B_s} C_{S}}{{m}_b} \Bigg|^2
+ \Bigg|
\frac{m_{B_s} C_{P}}{{m}_b} + \frac{2 m_\ell}{m_{B_s}} A_{10}\Bigg|^2
\Bigg\},
\end{eqnarray}
with $A_{10}$ defined in \eq{def:A10} and 
$C_{S,P}\equiv C_{S,P}(\mu)$, $m_b\equiv m_b (\mu)$.

\subsection{\bm$B \to K \ell^+ \ell^-$}
\begin{eqnarray}
\lefteqn{\frac{d \Gamma (B \to K \ell^+ \ell^-)}{d \hat{s}}  =  
  \frac{G_F^2  \aem^2  m_B^5}{2^{10} \pi^5} 
      \left| V_{tb}^{} V_{ts}^\ast  \right|^2  \hat{u}(\hat{s}) 
\Bigg\{ 
(|A^{\prime}|^2 +|C^{\prime}|^2) 
\Bigg[\lambda- \frac{\hat{u}(\hat{s})^2}{3} \Bigg]} \nnu\\
&+& 4 |C^{\prime}|^2  \hat{m}_\ell^2 (2+2 \hat{m}_K^2-\hat{s})
+ 8 \Re( C^{\prime} D^{\prime *}) \hat{m}_\ell^2 (1-\hat{m}_K^2)  
+4 |D^{\prime}|^2  \hat{m}_\ell^2 \hat{s} 
+ |T_P|^2 \hat{s} +  
|T_S|^2 (\hat{s}-4 \hat{m}_\ell^2) \nnu\\
&+&4 \Re(D^{\prime}T^*_P)  \hat{m}_\ell \hat{s} 
+4 \Re(C^{\prime}T^*_P)\hat{m}_\ell (1-\hat{m}_K^2) 
\Bigg\},
\label{eq:BKll}
\end{eqnarray}
with
\be
\label{eq:aux1}
  A^{\prime} =  \widetilde{C}_9^{\eff}(\hat{s}) f_+(\hat{s}) 
         + \frac{2 \hat{m}_b}{1 + \hat{m}_K} \widetilde{C}_7^{\eff} f_T(\hat{s}),
\ee
\be
C^{\prime}=  \widetilde{C}^\eff_{10}  f_+(\hat{s}),
\ee
\be
D^{\prime}  =   \frac{1-\hat{m}_K^2}{\hat{s}}\widetilde{C}^\eff_{10}[f_0(\hat{s}) - f_+(\hat{s})],
\ee
\be
T_{S,P} = \frac{1-\hat{m}_K^2}{\hat{m}_b} C_{S,P} f_0(\hat{s}) ,
\ee
where the definition of the form factors can be found in
\rf{Ali:1999mm}. The Wilson coefficients $\widetilde{C}_i^\eff$ can
be obtained from the ones in \eqs{eq:c7tilde}{eq:cis:NNLO} with $\omega_{7,9,79}=0$.

\subsection{\bm$B \to K^* \ell^+ \ell^-$}
\begin{eqnarray}
 \lefteqn{ \frac{d \Gamma(B \to K^* \ell^+ \ell^-)}{d \hat{s}}  =  
  \frac{G_F^2  \aem^2  m_B^5}{2^{10} \pi^5} 
      \left| V_{tb}^{} V_{ts}^\ast \right|^2  \hat{u}(\hat{s})
\Bigg\{ 
\frac{1}{3}\Bigg[|A|^2  \hat{s} \lambda \Bigg(1+\frac{2 \hat{m}_\ell^2}{\hat{s}}\Bigg)
+ |E|^2 \hat{s} \hat{u}(\hat{s})^2\Bigg]}  
              \nonumber \\
  &+& \frac{1}{4 \hat{m}_{K^*}^2} \Bigg( 
|B|^2 \Bigg[\lambda-\frac{\hat{u}(\hat{s})^2}{3} + 8 \hat{m}_{K^*}^2 (\hat{s} + 
2 \hat{m}_\ell^2) \Bigg] 
          + |F|^2 \Bigg[\lambda -\frac{ \hat{u}(\hat{s})^2}{3} + 
8 \hat{m}_{K^*}^2 (\hat{s}- 4 \hat{m}_\ell^2)\Bigg] 
\Bigg)
        \Bigg.
        \nonumber \\
  &+& 
   \frac{\lambda }{4 \hat{m}_{K^*}^2} \Bigg( |C|^2 \Bigg[\lambda - 
\frac{\hat{u} (\hat{s})^2}{3}\Bigg] 
 + |G|^2 \Bigg[\lambda -\frac{\hat{u}(\hat{s})^2}{3}+
4 \hat{m}_\ell^2(2+2 \hat{m}_{K^*}^2-\hat{s}) \Bigg] 
\Bigg)
             \nonumber \\
  &  -& 
   \frac{1}{2 \hat{m}_{K^*}^2}
\Bigg[ {\rm Re}(BC^\ast) \Bigg[\lambda -\frac{ \hat{u}(\hat{s})^2}{3}\Bigg]
(1 - \hat{m}_{K^*}^2 - \hat{s}) 
+    {\Re}(FG^\ast) \Bigg(\Bigg[\lambda -\frac{ \hat{u}(\hat{s})^2}{3}\Bigg]
(1 - \hat{m}_{K^*}^2 - \hat{s}) \nnu\\
&+& 4 \hat{m}_\ell^2 \lambda\Bigg) \Bigg] 
   - 2\hat{m}_\ell^2   [ {\Re}(FH^\ast)-
 {\Re}(GH^\ast)
 (1-\hat{m}_{K^*}^2)]\frac{\lambda}{\hat{m}_{K^*}^2} 
+\hat{m}_\ell^2   |H|^2\hat{s} \frac{\lambda }{\hat{m}_{K^*}^2} 
+  |X_P|^2 \hat{s} \frac{\lambda}{4 \hat{m}_{K^*}^2} \nnu\\
&+& 
|X_S|^2(\hat{s}-4 \hat{m}_\ell^2)\frac{\lambda}{4 \hat{m}_{K^*}^2}  -\hat{m}_\ell  
[\Re(F X_P^*)-(1-\hat{m}_{K^*}^2) \Re(G X_P^*)-\hat{s} 
\Re(H X_P^*) ]\frac{\lambda}{\hat{m}_{K^*}^2}
  \Bigg\}.
\end{eqnarray}
Here, 
\be
  A  =   \frac{2}{1 + \hat{m}_{K^*}} \widetilde{C}_9^{\eff}(\hat{s}) V(\hat{s}) 
         + \frac{4 \hat{m}_b}{\hat{s}}  \widetilde{C}_7^{\eff} T_1(\hat{s}),
\ee
\be
  B  =  (1 + \hat{m}_{K^*}) \Bigg[ \widetilde{C}_9^{\eff}(\hat{s}) A_1(\hat{s}) 
  + \frac{2 \hat{m}_b}{\hat{s}} (1 - \hat{m}_{K^*})  \widetilde{C}_7^{\eff} T_2(\hat
    s) \Bigg],
\ee
\be
  C  =  \frac{1}{1 - \hat{m}_{K^*}^2} \Bigg\{ 
         (1 - \hat{m}_{K^*}) \widetilde{C}_9^{\eff}(\hat{s}) A_2(\hat{s}) 
         + 2 \hat{m}_b  \widetilde{C}_7^{\eff} \Bigg[ 
  T_3(\hat{s}) + \frac{1 - \hat{m}_{K^*}^2}{\hat{s}} T_2(\hat{s}) \Bigg]
  \Bigg\},
\ee
\be
E  =  \frac{2}{1 + \hat{m}_{K^*}} \widetilde{C}^\eff_{10} V(\hat{s}),
\ee
\be
F =  (1 + \hat{m}_{K^*}) \widetilde{C}^\eff_{10} A_1(\hat{s}),
\ee
\be
  G  =  \frac{1}{1 + \hat{m}_{K^*}} \widetilde{C}^\eff_{10} A_2(\hat{s}),
\ee
\be
  H  =  \frac{1}{\hat{s}} \widetilde{C}^\eff_{10} \left[
 (1 + \hat{m}_{K^*}) A_1(\hat{s}) - (1 - \hat{m}_{K^*}) A_2(\hat{s}) - 
2 \hat{m}_{K^*} A_0(\hat{s}) \right],
\ee
\be
 X_{S,P}= - \frac{2\hat{m}_{K^*}}{\hat{m}_b} A_0 (\hat{s}) C_{S,P},
\ee
with the form factors defined in \rf{Ali:1999mm}. The
$\widetilde{C}_i^\eff$'s are given in 
Eqs.~(\ref{eq:c7tilde})-(\ref{eq:cis:NNLO}) with 
$\omega_{7,9,79}=0$.

\subsection{\bm$B \to X_s \ell^+ \ell^-$}

\begin{eqnarray}\label{diff:decay:rate:incl}
 \lefteqn{\frac{d \Gamma (B \to X_s \ell^+ \ell^-)}{d \hat{s}} = 
\frac{G_F^2   \aem^2 (m_{b}^\pole)^5 }{3 \times 2^{8} \pi^5} 
      \left| V_{tb}^{} V_{ts}^\ast  \right|^2   
(1-\hat{s})^2 \sqrt{1-\frac{4 \hat{m}_\ell^2}{\hat{s}^2}}}\nnu\\
&\times& \Bigg\{
\Bigg[12 \Re ( \widetilde{C}_7^{\eff} \widetilde{C}_9^{\eff\ast}) +
\frac{4 |\widetilde{C}_7^{\eff}|^2 (2+\hat{s})}{\hat{s}}\Bigg]
\Bigg(1+\frac{2 \hat{m}_\ell^2}{\hat{s}}\Bigg)
+ 6 \hat{m}_\ell^2(|\widetilde{C}_9^{\eff}|^2 - |\widetilde{C}_{10}^\eff|^2)\nnu\\
&+& (|\widetilde{C}_9^{\eff}|^2 + |\widetilde{C}_{10}^\eff|^2)\Bigg[1+2 \hat{s} +
  \frac{2\hat{m}_\ell^2 (1-\hat{s}) }{\hat{s}}\Bigg] 
+ \frac{3}{2} \hat{s}\Bigg[\Bigg(1-\frac{4
  \hat{m}_\ell^2}{\hat{s}}\Bigg) |C_S|^2 + |C_P|^2\Bigg]\nnu\\
&+& 6 \hat{m}_\ell \Re( C_P \widetilde{C}_{10}^{\eff \ast})\Bigg\},
\end{eqnarray}
with $\widetilde{C}_i^\eff$ defined in 
Eqs.~(\ref{eq:c7tilde})-(\ref{eq:cis:NNLO}). [Equation
(\ref{diff:decay:rate:incl}) agrees with \rf{Guetta:1997fw}  for
$m_s=0$.]

\newsection{Auxiliary coefficients}\label{recipe:Ai:Ci}
\bea
A_7 = \frac{4 \pi}{\alpha_s(\mu)} C_7(\mu)
- \frac{1}{3}  C_3(\mu)
- \frac{4}{9}  C_4(\mu)
- \frac{20}{3}  C_5(\mu)
- \frac{80}{9}  C_6(\mu),
\eea
\bea
A_8^{(0)} =  C^{(1)}_8(\mu)
+  C^{(0)}_3(\mu)
- \frac{1}{6}  C^{(0)}_4(\mu)
+ 20  C^{(0)}_5(\mu)
- \frac{10}{3}  C^{(0)}_6(\mu),
\eea
\bea
A_9 =  \frac{4 \pi}{\alpha_s(\mu)}  C_9(\mu) +
\sum_{i=1}^{6}  C_i(\mu)  \gamma_{i9}^{(0)}  \ln \frac{m_b}{\mu}
 + \frac{4}{3} C_3(\mu)
+ \frac{64}{9}  C_5(\mu)
+ \frac{64}{27}  C_6(\mu),
\eea
\bea\label{def:A10}
A_{10} =  \frac{4 \pi}{\alpha_s(\mu)} C_{10}(\mu),
\eea
\bea
T_9 =  \frac{4}{3} C_1(\mu) +  C_2(\mu)+ 6 C_3(\mu)+ 60 C_5(\mu) , 
\eea
\bea
U_9 =
- \frac{7}{2}  C_3(\mu) - \frac{2}{3}  C_4(\mu)
-38  C_5(\mu)
- \frac{32}{3} C_6(\mu),
\eea
\bea
W_9 =
- \frac{1}{2}  C_3(\mu) - \frac{2}{3} C_4(\mu)
-8  C_5(\mu)
- \frac{32}{3}  C_6(\mu),
\eea
where  $C_i(\mu) = C_i^{(0)}(\mu) + \alpha_s(\mu)/(4\pi)
C_i^{(1)}(\mu)+ \cdots$ and the $\gamma_{i9}^{(0)}$'s can be found in
\rf{NNLO:OB}.


\end{document}